\documentclass[10pt,twocolumn,journal]{IEEEtran}
\usepackage[T1]{fontenc}
\usepackage[latin9]{inputenc}
\usepackage{rotating}
\usepackage{float}
\usepackage{multirow}
\usepackage{amsmath}
\usepackage{amssymb}
\usepackage{graphicx}

\makeatletter

\newcommand{\noun}[1]{\textsc{#1}}
\providecommand{\tabularnewline}{\\}
\floatstyle{ruled}
\newfloat{algorithm}{tbp}{loa}
\providecommand{\algorithmname}{Algorithm}
\floatname{algorithm}{\protect\algorithmname}

\usepackage{amsfonts}
\usepackage{algorithm}% http://ctan.org/pkg/algorithms
\usepackage{algpseudocode}% http://ctan.org/pkg/algorithmicx
\makeatother

\usepackage{listings}

\begin{document}
\global\long\def\ie{\textit{i.e.,} }%
\global\long\def\eg{\textit{e.g.,} }%
\global\long\def\wrt{w.r.t.,}%

\title{Federated Learning with Cooperating Devices: \linebreak{}
 A Consensus Approach for Massive IoT Networks\thanks{\copyright2019 IEEE. Personal use of this material is permitted.
Permission from IEEE must be obtained for all other uses, in any current
or future media, including reprinting/republishing this material for
advertising or promotional purposes, creating new collective works,
for resale or redistribution to servers or lists, or reuse of any
copyrighted component of this work in other works. The authors are
with the Consiglio Nazionale delle Ricerche, Insitute of Electronics,
Computer and Telecommunication Engineering (IEIIT) and Politecnico
di Milano, DIG department. This work received support from the CHIST-ERA
III Grant RadioSense (Big Data and Process Modelling for the Smart
Industry - BDSI). The paper has been accepted for publication in the
IEEE Internet of Things Journal. The current arXiv contains an additional
Appendix C that describes the database and the Python scripts.}}
\author{Stefano Savazzi, \emph{Member, IEEE}, Monica Nicoli, \emph{Member,
IEEE}, Vittorio Rampa, \emph{Member, IEEE}}
\maketitle
\begin{abstract}
Federated learning (FL) is emerging as a new paradigm to train machine
learning models in distributed systems. Rather than sharing, and disclosing,
the training dataset with the server, the model parameters (\eg neural
networks weights and biases) are optimized collectively by large populations
of interconnected devices, acting as local learners. FL can be applied
to power-constrained IoT devices with slow and sporadic connections.
In addition, it does not need data to be exported to third parties,
preserving privacy. Despite these benefits, a main limit of existing
approaches is the centralized optimization which relies on a server
for aggregation and fusion of local parameters; this has the drawback
of a single point of failure and scaling issues for increasing network
size. The paper proposes a fully distributed (or server-less) learning
approach: the proposed FL algorithms leverage the cooperation of devices
that perform data operations inside the network by iterating local
computations and mutual interactions via consensus-based methods.
The approach lays the groundwork for integration of FL within 5G and
beyond networks characterized by decentralized connectivity and computing,
with intelligence distributed over the end-devices. The proposed methodology
is verified by experimental datasets collected inside an industrial
IoT environment. 
\end{abstract}

%\end{justify} 

%} 

\section{Introduction\label{sec:intro}}

\label{sec:Introduction}

Beyond 5G systems are expected to leverage cross-fertilizations between
wireless systems, core networking, Machine Learning (ML) and Artificial
Intelligence (AI) techniques, targeting not only communication and
networking tasks, but also augmented environmental perception services
\cite{iot2}. The combination of powerful AI tools, \emph{e.g.} Deep
Neural Networks (DNN), with massive usage of Internet of Things (IoT),
is expected to provide advanced services \cite{deep_surv} in several
domains such as Industry 4.0 \cite{computer}, Cyber-Physical Systems
(CPS) \cite{cps} and smart mobility \cite{vehicular}. Considering
this envisioned landscape, it is of paramount importance to integrate
emerging deep learning breakthroughs within future generation wireless
networks, characterized by arbitrary distributed connectivity patterns
(e.g., mesh, cooperative, peer-to-peer, or spontaneous), along with
strict constraints in terms of latency \cite{poor} (\ie to support
Ultra-Reliable Low-Latency Communications - URLLC) and battery lifetime.

\begin{figure}[!t]
\center\includegraphics[scale=0.49]{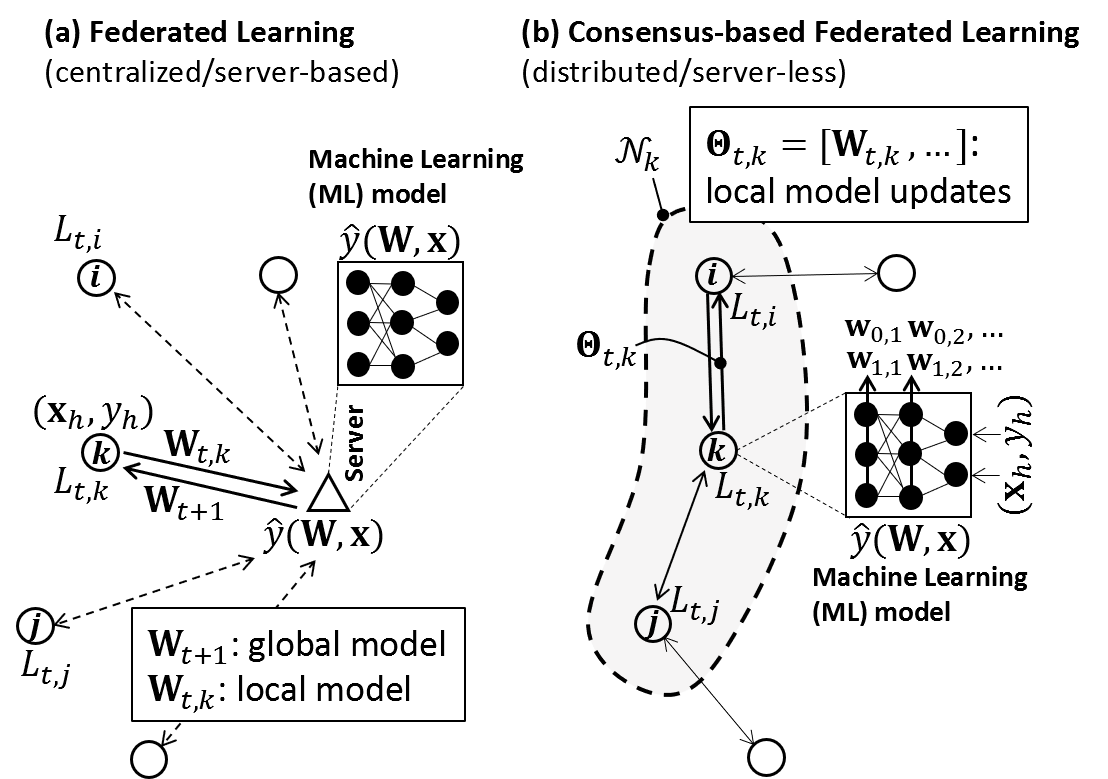} %\par\end{centering}
\protect\caption{\label{intro} From left to right: a) FL based on centralized fusion
of local model (or gradient) updates; b) proposed consensus-based
FL with distributed fusion over an infrastructure-less network. Learning
of global model parameters $\mathbf{W}$ from local data examples
$(\mathbf{x}_{h},y_{h})$ is obtained by mutual cooperation between
neighbors sharing the local model updates $\Theta_{t,k}$.}
\end{figure}

Recently, federated optimization, or federated learning (FL) \cite{kone1}\cite{feder}
has emerged as a new paradigm in distributed ML setups \cite{kone2}.
The goal of FL systems is to train a shared \emph{global model} (\emph{i.e.},
a Neural Network - NN) from a federation of participating devices
acting as local learners under the coordination of a central server
for models aggregation. As shown in Fig. \ref{intro}.a, FL alternates
between a local model computation at each device and a round of communication
with a server. Devices, or workers, derive a set of local learning
parameters from the available training data, referred to as \emph{local
model}. The local model $\mathbf{W}_{t,k}$ at time $t$ and device
$k$ is typically obtained via back-propagation and Stochastic Gradient
Descent (SGD) \cite{sgd} methods using local training examples $\textrm{(}i.e.,\textrm{data }\mathbf{x}_{h},\textrm{and labels }y_{h}\textrm{)}$.
The server obtains a global model by fusion of local models and then
feeds back such model to the devices. Multiple rounds are repeated
until convergence is reached. The objective of FL is thus to build
a global model $y=\hat{y}(\mathbf{W};\mathbf{x})$ by the cooperation
of a number of devices. Model is characterized by parameters $\mathbf{W}$
(\emph{i.e.}, NN weights and biases for each layer) for the output
quantity $y$ and the observed (input) data $\mathbf{x}$. Since it
decouples the ML stages from the need to send data to the server,
FL provides strong privacy advantages compared to conventional centralized
learning methods.

\subsection{Decentralized FL and related works}

Next generation networks are expected to be underpinned by new forms
of decentralized, infrastructure-less communication paradigms \cite{senseme}
enabling devices to cooperate directly over device-to-device (D2D)
spontaneous connections (\eg multihop or mesh). These networks are
designed to operate - when needed - without the support of a central
coordinator, or with limited support for synchronization and signalling.
They are typically deployed in mission-critical control applications
where edge nodes cannot rely on a remote unit for fast feedback and
have to manage part of the computing tasks locally \cite{mag}, cooperating
with neighbors to self disclose the information. Typical examples
are low-latency safety-related services for vehicular \cite{Brambilla}
or industrial \cite{cloud} applications. Considering these trends,
research activities are now focusing on fully decentralized (\ie
server-less) learning approaches. Optimization of the model running
on the devices with privacy constraints is also critical \cite{distillation}
for human-related applications.

To the authors knowledge, few attempts have been made to address the
problem of decentralized FL. In \cite{gossip-1}\cite{gossipgrad},
a gossip protocol is adopted for ML systems. Through sum-weight gossip,
local models are propagated over a peer-to-peer network. However,
in FL over D2D networks, gossip-based methods cannot be fully used
because of medium access control (MAC) and half-duplex constraints,
which are ignored in these early approaches. More recently, \cite{TORRENT}
considers an application of distributed learning for medical data
centers where several servers collaborate to learn a common model
over a fully connected network. However, network scalability/connectivity
issues are not considered at all. In \cite{gossip}, a segmented gossip
aggregation is proposed. The global model is split into non overlapping
subsets and local learners aggregate the segmentation sets from other
learners. Having split the model, by introducing ad-hoc subsets and
segment management tasks, the approach is extremely application-dependent
and not suitable for more general ML contexts. Finally, \cite{lalitha}\cite{fed}
propose a peer-to-peer Bayesian-like approach to iteratively estimate
the posterior model distribution. Focus is on convergence speed and
load balancing, yet simulations are limited to a few nodes, and the
proposed method cannot be easily generalized to NN systems trained
by incremental gradient (\ie SGD) or momentum based methods.

\subsection{Contributions}

This paper proposes the application of FL principles to massively
dense and fully decentralized networks that do not rely upon a central
server coordinating the learning process. As shown in Fig. \ref{intro}.b,
the proposed FL algorithms leverage the mutual cooperation of devices
that perform data operations inside the network (in-network) via consensus-based
methods \cite{gloria}. Devices independently perform training steps
on their local dataset (batches) based on a local objective function,
by using SGD and the fused models received from the neighbors. Next,
similarly to gossip \cite{gossip-1}, devices forward the model updates
to their one-hop neighborhood for a new consensus step. Unlike the
methods in \cite{gossip-1}-\cite{fed}, the proposed approach is
general enough to be seamlessly applied to any NN model trained by
SGD or momentum methods. In addition, in this paper we investigate
the scalability problem for varying NN model layer size, considering
large and dense D2D networks with different connectivity graphs. These
topics are discussed, for the first time, by focusing on an experimental
Industrial IoT (IIoT) setting.

The paper contributions are summarized in the following: 
\begin{itemize}
\item the federated averaging algorithm \cite{kone2}\cite{mah} is revisited
to allow local learners to implement consensus techniques by exchanging
local model updates: consensus also extends existing gossip approaches; 
\item a novel class of FL algorithms based on the iteratively exchange of
\emph{both} model updates \emph{and} gradients is proposed to improve
convergence and minimize the number of communication rounds, in exchange
for a more intensive use of D2D links and local computing; 
\item all presented algorithms are validated over large scale massive networks
with intermittent, sporadic or varying connectivity, focusing in particular
on an experimental IIoT setup, and considering both complexity, convergence
speed, communication overhead, and average execution time on embedded
devices. 
\end{itemize}
The paper is organized as follows. Sect. \ref{sec:Federated-optimization-and}
reviews the FL problem. Sect. \ref{sec:A-consensus-based-approach}
proposes two consensus-based algorithms for decentralized FL. Validation
of the proposed methods is first addressed in Sect. \ref{sec:Federated-averaging-by}
on a simple network and scenario. Then, in Sect. \ref{sec:Industrial-IoT:-experiments}
the validation is extended to a large scale setup by focusing on a
real-world problem in the IIoT domain. Finally, Sect. \ref{sec:Conclusions-and-open}
draws some conclusions and proposes future investigations.

\section{FL for model optimization}

\label{sec:Federated-optimization-and}

The FL approach defines an incremental algorithm for model optimization
over a large population of devices. It belongs to the family of incremental
gradient algorithms \cite{rabbat}\cite{LMS} but, unlike these setups,
optimization typically focuses on non-convex objectives that are commonly
found in NN problems. The goal is to learn a global model $\hat{y}(\mathbf{W};\mathbf{x})$
for inference problems (\ie classification or regression applications)
that transforms the input observation vector $\mathbf{x}$ into the
outputs $\hat{y}\in\left\{ y_{c}\right\} _{c=1}^{C}$, with model
parameters embodied in the matrix $\mathbf{W}$ while $C$ is the
output size. Observations, or input data, are stored across $K$ devices
connected with a central server that coordinates the learning process.
A common, and practical, assumption is that the number of participating
devices $K$ is large ($K\gg1$) and they have intermittent connectivity
with the server. The cost of communication is also much higher than
local computation, in terms of capacity, latency and energy consumption
\cite{disco}. In this paper, we will focus specifically on NN models.
Therefore, considering a NN of $Q\geq1$ layers, the model iteratively
computes a non-linear function $f(\cdot)$ of a weighted sum of the
input values, namely 
\begin{equation}
\hat{y}(\mathbf{W};\mathbf{x})=f_{Q}\left(\mathbf{w}_{0,Q}^{\mathrm{T}}\mathbf{h}_{Q-1}+\mathbf{w}_{1,Q}\right)\label{eq:ff}
\end{equation}
with $\mathbf{h}_{q}=f_{q}\left(\mathbf{w}_{0,q}^{\mathrm{T}}\mathbf{h}_{q-1}+\mathbf{w}_{1,q}\right)$,
$\ensuremath{q=1,...,Q-1}$, being the hidden layers and $\mathbf{h}_{0}=\mathbf{x}$
the input vector. The matrix 
\begin{equation}
\mathbf{W}=\left[\mathbf{w}_{0,1}^{\mathrm{T}},\mathbf{w}_{1,1},...,\mathbf{w}_{0,Q}^{\mathrm{T}},\mathbf{w}_{1,Q}\right]\label{eq:mat}
\end{equation}
collects all the parameters of the model, namely the weights $\mathbf{w}_{0,q}\in\mathbb{R}^{d_{1}\times d_{2}}$
and the biases $\mathbf{w}_{1,\mathit{q}}\in\mathbb{R}^{d_{2}\times1}$
for each defined layer, with $d_{1}$ and $d_{2}$ the corresponding
input and output layer dimensions, respectively\footnote{To simplify the notation, here we assume that the layers have equal
input/output size in $\mathbf{W}$, but the model can be generalized
to account for different dimensions (see Sect. V).}. FL applies \emph{independently} to each layer\footnote{Weights and biases are also optimized independently.}
of the network. Therefore, in what follows, optimization focuses on
one layer $q$ and the matrix (\ref{eq:mat}) reduces to $\mathbf{W=\left[\text{\ensuremath{\mathbf{w}_{0,q}^{\mathrm{T}}},\ensuremath{\mathbf{w}_{1,q}}}\right]}$.
The parameters of the convolutional layers, namely input, output and
kernel dimensions, can be easily reshaped to conform with the above
general representation.

In FL it is commonly assumed \cite{kone1}\cite{feder} that a large
database of examples is (unevenly) partitioned among the $K$ devices,
under non Identical Independent Distribution (non-IID) assumptions.
Examples are thus organized as the tuples $(\mathbf{x}_{h},y_{h})$,
$h=1,...,E$ where $\mathbf{x}_{h}$ represents the data, while $y_{h}$
are the desired model outputs $\hat{y}$. The set of examples, or
training data, available at device $k$ is $\mathcal{E}_{k}$, where
$E_{k}=\left|\mathcal{E}_{k}\right|\ll E$ is the size of the \emph{k}-th
dataset under the non-IID assumption. The training data on a given
device is thus not representative of the full population distribution.
In practical setups (see Sect. \ref{sec:Industrial-IoT:-experiments}),
data is collected individually by the devices based on their local/partial
observations of the given phenomenon.

Unlike incremental gradient algorithms \cite{diffusion-1}\cite{diffusion},
FL of model $\mathbf{W}$ is applicable to any finite-sum objective
$L(\mathbf{W})$ of the general form

\begin{equation}
\underset{\mathbf{W}}{\mathrm{min}}\thinspace L(\mathbf{W})=\underset{\mathbf{W}}{\mathrm{min}}\underset{L(\mathbf{W})}{\underbrace{\sum_{k=1}^{K}\frac{E_{k}}{E}\times L_{k}(\mathbf{W})},}\label{eq:fdd}
\end{equation}
where $L_{k}(\mathbf{W})$ is the loss, or cost, associated with the
$k$-th device

\begin{equation}
L_{k}(\mathbf{W})=\frac{1}{E_{k}}\sum_{h=1}^{E_{k}}\ell(\mathbf{x}_{h},y_{h};\mathbf{W})\label{eq:lossk}
\end{equation}
and $\ell(\mathbf{x}_{h},y_{h};\mathbf{W})$ is the loss of the predicted
model over the $E_{k}$ examples $(\mathbf{x}_{h},y_{h})$ observed
by the device $k$, assuming model parameters $\mathbf{W}$ to hold.

In conventional centralized ML (\ie learning \emph{without} federation),
used here as benchmark, the server collects \emph{all} local training
data from the devices and obtains the optimization of model parameters
by applying an incremental gradient method over a number of batches
from the training dataset. For iteration $t$, the model parameters
are thus updated by the server according to 
\begin{equation}
\mathbf{W}_{t+1}=\mathbf{W}_{t}-\mu_{s}\times\nabla L(\mathbf{W}_{t}),\label{nofederation}
\end{equation}
where $\mu_{s}$ is the SGD step size and $\nabla L(\mathbf{W}_{t})=\nabla_{\mathbf{W}_{t}}\left[L(\mathbf{W}_{t})\right]$
the gradient of the loss in (\ref{eq:fdd}) over the assigned batches
and \emph{w.r.t.} the model $\mathbf{W}_{t}$. Backpropagation is
used here for gradients computation. The model estimate at convergence
is denoted as $\mathbf{W}_{\infty}=\lim_{t\rightarrow\infty}\mathbf{W}_{t}$.

Rather than sharing the training data with the server, in FL the model
parameters $\mathbf{W}$ are optimized collectively by interconnected
devices, acting as local learners. On each round $t$ of communication,
the server distributes the current global model $\mathbf{W}_{t}$
to a subset $\mathcal{S}_{t}$ of $n_{t}$ devices. The devices independently
update the model $\mathbf{W}_{t}$ using the gradients (SGD) from
\emph{local} training data as 
\begin{equation}
\mathbf{W}_{t+1,k}=\mathbf{W}_{t}-\mu\times\nabla L_{t,k}(\mathbf{W}_{t}),\label{eq:update}
\end{equation}
where $\nabla L_{t,k}(\mathbf{W}_{t})=\nabla_{\mathbf{W}_{t}}\left[L_{t,k}(\mathbf{W}_{t})\right]$
represents the gradient of the loss (\ref{eq:lossk}) observed by
the $k$-th device and \emph{w.r.t.} the model $\mathbf{W}_{t}$.
Updates $\nabla L_{t,k}$ (\ref{eq:update}), or local models $\mathbf{W}_{t+1,k}$,
are sent back to the server, after quantization, anonymization \cite{kone1}
and compression stages, modelled here by the operator $\mathcal{P}_{\Theta}$.
A global model update is obtained by the server through aggregation
according to 
\begin{equation}
\mathbf{W}_{t+1}=\mathbf{W}_{t}-\mu_{s}\frac{1}{n_{t}}\sum_{k=1}^{n_{t}}\frac{E_{k}}{E}\mathcal{P}_{\Theta}\left[\nabla L_{t,k}(\mathbf{W}_{t})\right].\label{eq:learn}
\end{equation}
Convergence towards the centralized approach (\ref{nofederation})
is achieved if $\lim_{t\rightarrow\infty}\mathbf{W}_{t}=\mathbf{W}_{\infty}$.
Notice that the learning rate $\mu_{s}$ is typically kept smaller
\cite{kone2} compared with centralized learning (\ref{nofederation})
on large datasets. Aggregation model (\ref{eq:learn}) is referred
to as \emph{Federated Averaging} (FA) \cite{kone1}\cite{mah}. As
far as convergence is concerned, for strongly convex objective $L(\mathbf{W})$
and generic local solvers, the general upper bound on global iteration
number $N_{I}$ is given in \cite{Ma} and relates both to global
($\gamma_{G}$) and local ($\gamma_{L}$) accuracy according to the
equation $N_{I}=\mathcal{O\left(\log\left[\mathrm{1/\left(1-\gamma_{G}\right)}\right]\mathrm{/\gamma_{L}}\right)}$.

\section{A consensus-based approach to in-network FL}

\label{sec:A-consensus-based-approach}

The approaches proposed in this section allow the devices to learn
the model parameters, solution of (\ref{eq:fdd}), by relying \emph{solely}
on local cooperation with neighbors, and local in-network (as opposed
to centralized) processing. The interaction topology of the network
is modelled as a directed graph $\mathcal{G}=(\mathcal{V},\xi)$ with
the set of nodes $\mathcal{V}=\left\{ 1,2,...,K\right\} $ and edges
(links) $\xi$. As depicted in Fig. \ref{intro}, the $K$ distributed
devices are connected through a decentralized communication architecture
based on D2D communications. The neighbor set of device $k$ is denoted
as $\mathcal{N}_{k}=\left\{ i\in V:(i,k)\in\xi\right\} $, with cardinality
$\left|\mathcal{N}_{k}\right|$. Notice that we include node $k$
in the set $\mathcal{N}_{k}$, while $\mathcal{N}_{\bar{k}}=\mathcal{N}_{k}\backslash\left\{ k\right\} $
does not. As introduced in the previous section, each device has a
database $\mathcal{E}_{k}$ of examples $(\mathbf{x}_{h},y_{h})$
that are used to train a local NN model $\mathbf{W}_{t,k}$ at some
time $t$ (epoch). The model maps input features $\mathbf{x}$ into
outputs $\hat{y}(\mathbf{W}_{t,k};\mathbf{x})$ as in (\ref{eq:ff}).
A cost function, generally non-convex, as $L_{k}(\mathbf{W}_{t,k})$
in (\ref{eq:lossk}), is used to optimize the weights $\mathbf{W}_{t,k}$
of the local model.

The proposed FL approaches exploit both adaptive diffusion \cite{diffusion}
and consensus tools \cite{gloria}\cite{consensus} to optimally leverage
the (possibly large) population of federated devices that cooperate
for the distributed estimation of the global model $\mathbf{W}$,
while retaining the trained data. Convergence is thus obtained if
$\forall k$ it is $\lim_{t\rightarrow\infty}\mathbf{W}_{t,k}=\mathbf{W}_{\infty}$.
Distributed in-network model optimization must satisfy convergence
time constraints, as well as minimize the number of communication
rounds. In what follows, we propose two strategies that differ in
the way the model updates $\mathbf{W}_{t,k}$ and gradients $\nabla L_{t,k}$
are computed and updated.

\begin{algorithm}[t]
\caption{Consensus-based Federated Averaging}

\label{cfa} \begin{algorithmic}[1]

\Procedure{CFA}{$\mathcal{N}_{\bar{k}},\epsilon_{t},\alpha_{t,i}$}

\State initialize $\mathbf{W}_{0,k}$ $\gets$ device $k$

\For{each round $t=1,2,...$}\Comment{Main loop}

\State$\mathbf{receive}$$\left\{ \mathbf{W}_{t,i}\right\} _{i\in\mathcal{N}_{\bar{k}}}$\Comment{RX
from neighbors}

\State$\boldsymbol{\psi}_{t,k}\gets\mathbf{W}_{t,k}$

\For{all devices $i\in\mathcal{N}_{\bar{k}}$}

\State$\boldsymbol{\psi}_{t,k}\gets\boldsymbol{\psi}_{t,k}+\epsilon_{t}\alpha_{t,i}\left(\mathbf{W}_{t,i}-\mathbf{W}_{t,k}\right)$

\EndFor

\State$\mathbf{W}_{t+1,k}=\mathrm{\text{ModelUpdate}}(\mathit{\boldsymbol{\psi}_{t,k}})$

\State$\mathbf{send}$($\mathbf{W}_{t+1,k}$)\Comment{TX to neighbors}

\EndFor

\EndProcedure

\Procedure{ModelUpdate}{$\mathit{\boldsymbol{\psi}_{t,k}}$}\Comment{Local
SGD}

\State$\mathcal{B}\gets$ mini-batches of size $B$

\For{batch \texttt{\textbf{$b\in\mathcal{B}$}}}\Comment{Local
model update}

\State $\boldsymbol{\psi}_{t,k}\gets\mathit{\boldsymbol{\psi}_{t,k}}-\mu_{t}\nabla L_{t,k}(\boldsymbol{\psi}_{t,k})$

\EndFor

\State$\mathbf{W}_{t,k}\gets\boldsymbol{\psi}_{t,k}$

\State return($\mathbf{W}_{t,k}$)

\EndProcedure

\end{algorithmic} 
\end{algorithm}

\subsection{Consensus based Federated Averaging (CFA)}

The first strategy extends the centralized FA and it is described
in the pseudocode fragment of Algorithm \ref{cfa}. It is referred
to as Consensus-based Federated Averaging (CFA).

After initialization\footnote{Each device hosts a model $\mathbf{W}$ of the same architecture and
initialized similarly.} of $\mathbf{W}_{0,k}$ at time $t=0$, on each communication round
$t>0$, device $k$ sends its model updates $\mathbf{W}_{t,k}$ (once
per round) and receives weights from neighbors $\mathbf{W}_{t,i}$,
$i\in\mathcal{N}_{\bar{k}}$. Based on received data, the device updates
its model $\mathbf{W}_{t,k}$ sequentially to obtain the aggregated
model 
\begin{equation}
\boldsymbol{\psi}_{t,k}=\mathbf{W}_{t,k}+\epsilon_{t}\sum_{i\in\mathcal{N}_{\bar{k}}}\alpha_{k,i}\left(\mathbf{W}_{t,i}-\mathbf{W}_{t,k}\right),\label{eq:aggreg}
\end{equation}
where $\epsilon_{t}$ is the \emph{consensus step-size} and $\alpha_{k,i}$,
$i\in\mathcal{N}_{\bar{k}}$, are the mixing weights for the models.
Next, gradient update is performed using the aggregated model $\boldsymbol{\psi}_{t,k}$
as 
\begin{equation}
\mathbf{W}_{t+1,k}=\boldsymbol{\psi}_{t,k}-\mu_{t}\nabla L_{t,k}(\boldsymbol{\psi}_{t,k}),\label{adapt}
\end{equation}
by running SGD over a number of mini-batches of size $B<E_{k}$. Model
aggregation (\ref{eq:aggreg}) is similar to the sum-weight gossip
protocol \cite{gossip-1,gossipgrad}, when setting $\epsilon_{t}=1$.
However, mixing weights $\alpha_{k,i}$ are used here to combine model
innovations, $\left\{ \mathbf{W}_{t,i}-\mathbf{W}_{t,k}\right\} $,
$i\in\mathcal{N}_{\bar{k}}$. In addition, the step-size $\epsilon_{t}$
controls the consensus stability.

Inspired by FA approaches (Sect. \ref{sec:Federated-optimization-and}),
the mixing weights $\alpha_{k,i}$ are chosen as 
\begin{equation}
\alpha_{k,i}=\frac{E_{i}}{\sum_{i\in\mathcal{N}_{\bar{k}}}E_{i}}.\label{eq:mix}
\end{equation}
Other choices are based on weighted consensus strategies \cite{gloria},
where the mixing weights $\alpha_{k,i}$ are adapted on each epoch
$t$ based on current validation accuracy or loss metrics. The consensus
step-size $\epsilon_{t}$ can be chosen as $\epsilon_{t}\in\left(0,1/\Delta\right),$
where $\Delta=\max_{k}\left(\sum_{i\in\mathcal{N}_{\bar{k}}}\alpha_{k,i}\right)$,
namely the maximum degree of the graph $\mathcal{G}$ \cite{saber}
that models the interaction topology of the network. Notice that the
graph $\mathcal{G}$ has adjacency matrix $\mathit{\mathbf{A}}=[a_{k,i}]$
where $a_{k,i}=\alpha_{k,i}$ iff $i\in\mathcal{N}_{\bar{k}}$ and
$a_{k,i}=0$ otherwise. Beside consensus step-size, it is additionally
assumed that the SGD step-size $\mu_{t}$ is optimized for convergence:
namely, the objective function value $L_{t,k}$ is decreasing with
each iteration of gradient descent, or after some threshold. Convergence
is further analyzed in Sect. \ref{sec:Industrial-IoT:-experiments}
with experimental data.

By defining as $\Theta_{t,k}$ the set of parameters to be exchanged
among neighbors, CFA requires the iterative exchange of model updates
$\mathbf{W}_{t,i},$ $\forall i\in\mathcal{N}_{k}$, therefore 
\begin{equation}
\Theta_{t,k}:=\left[\mathbf{W}_{t,k}\right].\label{eq:paramcfa}
\end{equation}

\subsection{Consensus based Federated Averaging with Gradients Exchange (CFA-GE)}

\label{subsec:Consensus-based-Federated}

\begin{figure}[t]
\center\includegraphics[scale=0.31]{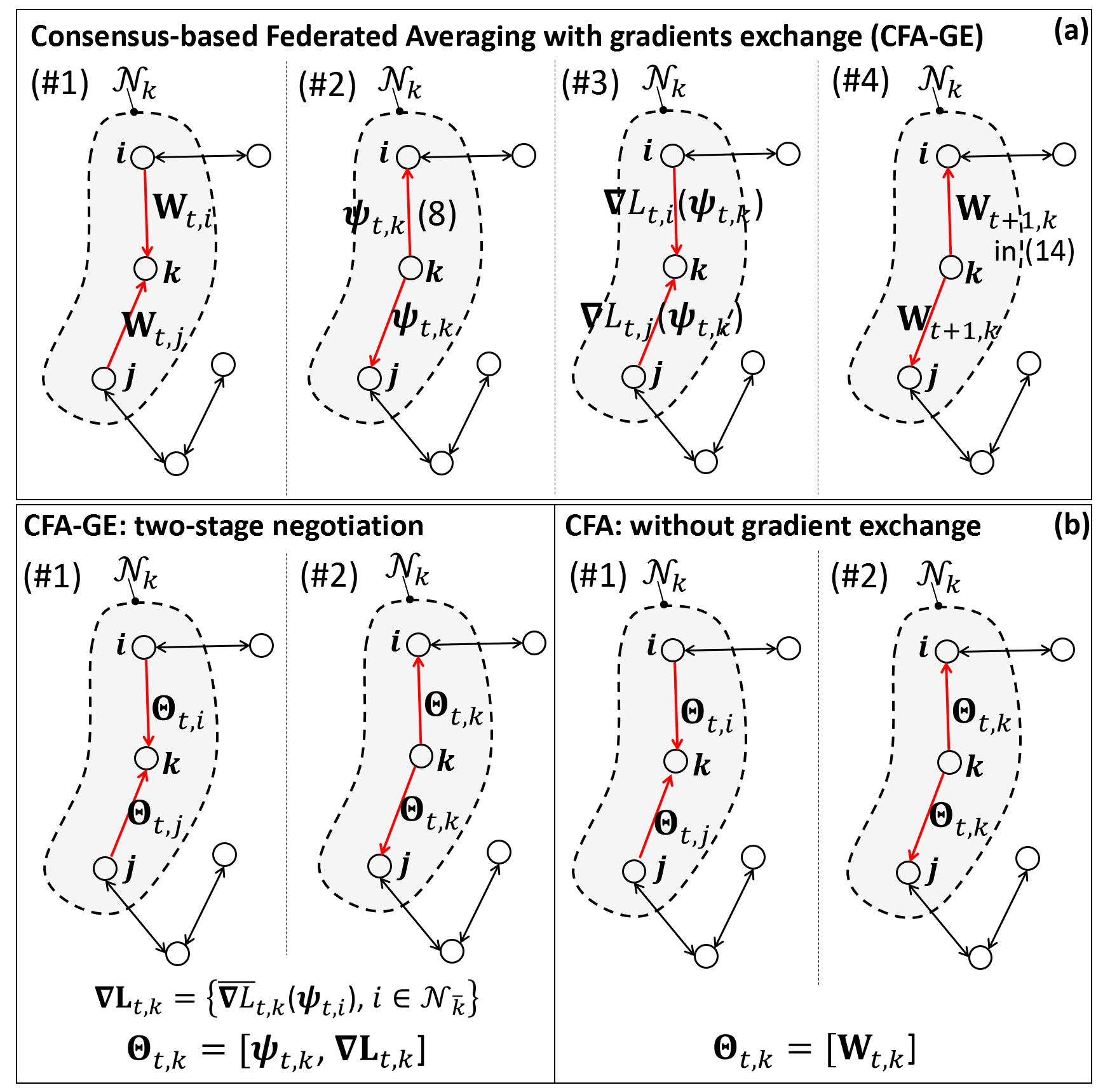} %\par\end{centering}
\protect\caption{\label{exchange} From top to bottom: a) CFA-GE; b) CFA-GE with two-stage
negotiation and implementation (left) compared against CFA w/o gradient
exchange (right). }
\end{figure}

The second strategy proposes the \emph{joint} exchange of local gradients
\emph{and} model updates by following the four-stage iterative procedure
illustrated in the Fig. \ref{exchange}.a for epoch $t$. The new
algorithm is referred to as Consensus-based Federated Averaging with
Gradients Exchange (CFA-GE). The first stage (step \#1) is similar
to CFA and obtains $\boldsymbol{\psi}_{t,k}$ by consensus-based model
aggregation in (\ref{eq:aggreg}). Before using $\boldsymbol{\psi}_{t,k}$
for the local model update, it is fed back to the same neighbors (``negotiation''
stage in step \#2 of Fig. \ref{exchange}.b). Model $\boldsymbol{\psi}_{t,k}$
is then used by the neighbors to compute the gradients 
\begin{equation}
\nabla L_{t,i}(\boldsymbol{\psi}_{t,k}),\:\forall i\in\mathcal{N}_{\bar{k}}\label{eq:gradients}
\end{equation}
using \emph{their} local data. Notice that all gradients are computed
over a single batch\footnote{Sending multiple gradients (corresponding to mini-batches) is an alternative
option, not considered here for bandwidth limitations. } (or mini-batch) of local data, while the chosen batch/mini-batch
can change on consecutive communication rounds. Gradients are sent
back to the device $k$ in step \#3. Compared with CFA, this step
allows every device to exploit additional gradients using neighbor
data, and makes the learning much faster. On the device $k$, the
local model is thus updated using the \emph{received }gradients (\ref{eq:gradients})
according to

\begin{equation}
\widetilde{\boldsymbol{\psi}}_{t,k}=\boldsymbol{\psi}_{t,k}-\mu_{t}\sum_{i\in\mathcal{N}_{\bar{k}}}\beta_{k,i}\mathcal{P}_{\Theta}\left[\nabla L_{t,i}(\boldsymbol{\psi}_{t,k})\right],\label{adapt-1-1}
\end{equation}
where $\beta_{k,i}$ are the mixing weights for the gradients. Finally,
as done for CFA in (\ref{adapt}), the gradient update is performed
using now the aggregated model $\widetilde{\boldsymbol{\psi}}_{t,k}$
(\ref{adapt-1-1}) and \emph{local} data mini-batches

\begin{equation}
\mathbf{W}_{t+1,k}=\widetilde{\boldsymbol{\psi}}_{t,k}-\mu_{t}\nabla L_{t,k}(\widetilde{\boldsymbol{\psi}}_{t,k}).\label{adapt-1}
\end{equation}
To summarize, for each device $k$, CFA-GE combines the gradients
$\nabla L_{t,k}(\boldsymbol{\psi}_{t,k})$ computed over the local
data with the gradients $\nabla L_{t,i}(\boldsymbol{\psi}_{t,k})$,
$i\in\mathcal{N}_{\bar{k}}$ obtained by the neighbors over their
batches. The negotiation stage (\ref{adapt-1-1})-(\ref{adapt-1})
is similar to the diffusion strategy proposed in \cite{diffusion-1}\cite{diffusion}.
In particular, we aggregate the model first (\ref{eq:aggreg}), then
we run one gradient descent round using the received gradients (\ref{adapt-1-1}),
and finally, a number of SGD rounds (\ref{adapt-1}) using local mini-batches.
As revealed in Sect. \ref{sec:Industrial-IoT:-experiments}, optimization
of the mixing weights $\beta_{k,i}$ for the gradients is critical
for convergence. Considering that the gradients in (\ref{eq:gradients}),
obtained from neighbors, are computed over a single batch of data,
as opposed to local data mini-batches, a reasonable choice is $\beta_{k,i}>1$,
$\forall$$i\in\mathcal{N}_{\bar{k}}$. This aspect is further discussed
in Sect. \ref{sec:Industrial-IoT:-experiments}.

\begin{algorithm}[t]
\caption{CFA with gradients exchange }

\label{CFA-ge}\begin{algorithmic}[1]

\Procedure{CFA-GE}{$\mathcal{N}_{\bar{k}},\epsilon_{t},\alpha_{t,i},\beta_{t,i}$}

\State initialize $\mathbf{W}_{0,k},\boldsymbol{\psi}_{1,k},\nabla L_{1,k}(\boldsymbol{\psi}_{0,k})$

\For{each round $t=2,...$}\Comment{Main loop}

\State$\mathbf{receive}$$\left\{ \boldsymbol{\psi}_{t,i},\overline{\nabla L}_{t,i}(\boldsymbol{\psi}_{t-1,k})\right\} _{i\in\mathcal{N}_{\bar{k}}}$\Comment{RX}

\State$\boldsymbol{\psi}_{t,k}\gets\mathbf{W}_{t,k}$

\For{all devices $i\in\mathcal{N}_{\bar{k}}$}

\State$\boldsymbol{\psi}_{t,k}\gets\boldsymbol{\psi}_{t,k}+\epsilon_{t}\alpha_{t,i}\left(\boldsymbol{\psi}_{t,i}-\mathbf{W}_{t,k}\right)$

\State$\mathbf{compute}$ $\nabla L_{t+1,k}(\boldsymbol{\psi}_{t,i})$\Comment{gradients}

\State$\overline{\nabla L}_{t+1,k}\leftarrow$ in $(\ref{eq:grad_moving})$\Comment{MEWMA
update}

\EndFor

\State$\boldsymbol{\widetilde{\psi}}_{t,k}\gets\boldsymbol{\psi}_{t,k}$

\For{all devices $i\in\mathcal{N}_{\bar{k}}$}

\State$\boldsymbol{\widetilde{\psi}}_{t,k}\negmedspace\leftarrow\negmedspace\boldsymbol{\widetilde{\psi}}_{t,k}\negthinspace-\negthinspace\mu_{t}\beta_{t,i}\overline{\nabla L}_{t,i}(\boldsymbol{\psi}_{t-1,k})$

\EndFor

\State$\mathbf{W}_{t+1,k}\leftarrow\mathrm{ModelUpdate}\mathit{(\boldsymbol{\widetilde{\psi}}_{t,k})}$

\State$\boldsymbol{\psi}_{t+1,k}\leftarrow\boldsymbol{\psi}_{t,k}$

\State$\nabla\mathbf{L}_{t+1,k}:=\left\{ \overline{\nabla L}_{t+1,k},\text{}\forall i\in\mathcal{N}_{\bar{k}}\right\} .$

\State$\mathbf{send}$($\boldsymbol{\psi}_{t+1,k},\nabla\mathbf{L}_{t+1,k}$)\Comment{TX
to neighbors}

\EndFor

\EndProcedure

\end{algorithmic} 
\end{algorithm}

\subsection{Two-stage negotiation and implementation aspects}

\label{subsec:Two-stage-negotiation-and}

Unlike CFA, CFA-GE requires a larger use of the bandwidth and more
communication rounds for the synchronous exchange of the gradients.
More specifically, it requires a more intensive use of the D2D wireless
links for sharing models first during the negotiations (step \#2)
and then forwarding gradients (step \#3). In addition, each device
should wait for neighbor gradients before applying any model update.
Here, the proposed implementation simplifies the negotiation procedure
to improve convergence time (and latency). In particular, it resorts
to a two-stage scheme, while, likewise CFA, each device can perform
the updates without waiting for a reply from neighbors. Pseudocode
is highlighted in Algorithm \ref{CFA-ge}. Communication rounds vs.
epoch $t$ for CFA-GE are detailed in Fig. \ref{exchange}.b and compared
with CFA. Considering the device $k$, with straightforward generalization,
the following parameters are exchanged with neighbors at epoch $t$
as 
\begin{equation}
\Theta_{t,k}:=\left[\boldsymbol{\psi}_{t,k},\nabla\mathbf{L}_{t,k}\right],\label{eq:param}
\end{equation}
namely the model updates (aggregations) $\boldsymbol{\psi}_{t,k}$
and the gradients $\nabla\mathbf{L}_{t,k}$, organized as 
\begin{equation}
\nabla\mathbf{L}_{t,k}:=\left\{ \overline{\nabla L}_{t,k}(\boldsymbol{\psi}_{t-1,i}),\text{}\forall i\in\mathcal{N}_{\bar{k}}\right\} .\label{eq:gradients_matrix}
\end{equation}
In the proposed two stage implementation, the negotiation step (step
\#2 in Fig. \ref{exchange}.a) is not implemented as it requires a
synchronous model sharing. Therefore, $\forall i\in\mathcal{N}_{\bar{k}}$
the model aggregations $\boldsymbol{\psi}_{t,i}$ are not available
by device $k$ at epoch $t$, or, equivalently, the device $k$ does
not wait for such information from the neighbors. The gradients $\nabla L_{t,k}(\boldsymbol{\psi}_{t,i})$
are now \emph{predicted} as $\overline{\nabla L}_{t,k}(\boldsymbol{\psi}_{t-1,i})$
using the past (outdated) models $\boldsymbol{\psi}_{t-1,i}$,$\boldsymbol{\psi}_{t-2,i},...$
from the neighbors. In line with momentum based techniques (see \cite{momentum}
and also Appendix B), for the predictions $\overline{\nabla L}_{t,k}$
we use a multivariate exponentially weighted moving average (MEWMA)
of the past gradients 
\begin{equation}
\overline{\nabla L}_{t,k}(\boldsymbol{\psi}_{t-1,i})=\varrho\nabla L_{t,k}(\boldsymbol{\psi}_{t-1,i})+(1-\varrho)\overline{\nabla L}_{t-1,k}\label{eq:grad_moving}
\end{equation}
ruled by the hyper-parameter $0<\varrho\leq1$. Setting $\varrho=1$,
the gradient is estimated using the last available model ($\boldsymbol{\psi}_{t-1,i}$):
$\overline{\nabla L}_{t,k}(\boldsymbol{\psi}_{t-1,i})=\nabla L_{t,k}(\boldsymbol{\psi}_{t-1,i})$.
A smaller value $\varrho<1$ introduces a memory $(1-\varrho)\overline{\nabla L}_{t-1,k}$
with $\overline{\nabla L}_{t-1,k}=\overline{\nabla L}_{t-1,k}(\boldsymbol{\psi}_{t-2,i},...)$
depending on the past models $\boldsymbol{\psi}_{t-2,i},...,$. This
is shown, in Sect. \ref{sec:Industrial-IoT:-experiments}, to be beneficial
on real data.

Assuming that the device $k$ is able to correctly receive and decode
the messages from the neighbors $\Theta_{t,i}=\left[\boldsymbol{\psi}_{t,i},\nabla\mathbf{L}_{t,i}\right]$,
$\forall i\in\mathcal{N}_{\bar{k}}$ at epoch $t$, the model aggregation
step changes from (\ref{eq:aggreg}) to

\begin{equation}
\boldsymbol{\psi}_{t,k}=\mathbf{W}_{t,k}+\epsilon_{t}\sum_{i\in\mathcal{N}_{\bar{k}}}\alpha_{k,i}\left(\boldsymbol{\psi}_{t,i}-\mathbf{W}_{t,k}\right),\label{eq:aggreg-fast}
\end{equation}
while the model update step using the \emph{received }gradients is
now

\begin{equation}
\widetilde{\boldsymbol{\psi}}_{t,k}=\boldsymbol{\psi}_{t,k}-\mu_{t}\sum_{i\in\mathcal{N}_{\bar{k}}}\beta_{k,i}\mathcal{P}_{\Theta}\left[\overline{\nabla L}_{t,i}(\boldsymbol{\psi}_{t-1,k})\right]\label{eq:adapt-fast}
\end{equation}
and replaces (\ref{adapt-1-1}). Finally, a gradient update on local
data is done as in (\ref{adapt-1}). Notice that Algorithm \ref{CFA-ge}
implements (\ref{eq:adapt-fast}) by running one gradient descent
round per received gradient (lines $12$-$14$) to allow for asynchronous
updates. In the Appendix B, we discuss the application of CFA and
CFA-GE to advanced SGD strategies designed to leverage momentum information
\cite{sgd}\cite{momentum}.

\subsection{Communication overhead and complexity analysis}

\label{subsec:Communication-overhead-and}

With respect to FA, the proposed decentralized algorithms take some
load off of the server, at the cost of additional in-network operations
and increased D2D communication overhead. Overhead is quantified here
for both CFA and CFA-GE in terms of the size of the parameters $\Theta_{t,k}$
that need to be exchanged among neighbors. CFA extends FA and, similarly,
requires each node to exchange only local model updates at most once
per round. The overhead, or the size of $\Theta_{t,k}$, thus corresponds
to the model size (\ref{eq:paramcfa}). For a generic DNN model of
$Q$ layers, the model $\mathbf{W}_{t,k}$ size can be approximated
in the order of $\mathcal{O}(d_{1}d_{2}Q)\ll E$. This is several
order of magnitude lower than the size of the input training data,
in typical ML and deep ML problems. As in (\ref{eq:param}), CFA-GE
requires the exchange of local model aggregations $\boldsymbol{\psi}_{t,k}$
and one gradient $\overline{\nabla L}_{t,k}(\boldsymbol{\psi}_{t-1,i})$
per neighbor, $\forall i\in\mathcal{N}_{\bar{k}}$. Overhead now scales
with $\mathcal{O}(d_{1}d_{2}Q\left|\mathcal{N}_{k}\right|)$, where
$\left|\mathcal{N}_{k}\right|=\left|\mathcal{N}_{\bar{k}}\right|+1$.
This is still considerably lower than the training dataset size, provided
that the number of participating neighbors is limited. In the examples
of Sect. \ref{sec:Industrial-IoT:-experiments}, we show that $\left|\mathcal{N}_{\bar{k}}\right|=2$
neighbors are sufficient, in practice, to achieve convergence: notice
that the number of active neighbors is also typically small to avoid
traffic issues \cite{M2M}. Finally, quantization $\mathcal{P}_{\Theta}\left[\cdot\right]$
of the parameters can be also applied to limit the transmission payload,
with the side effect to improve also global model generalization \cite{icassp-compr}.

Besides overhead, CFA and CFA-GE computational complexity scales with
the global model size and it is ruled by the number of local SGD rounds.
However, unlike FA, model aggregations \emph{and} local SGD are both
implemented on the device. With respect to CFA, CFA-GE computes up
to $\left|\mathcal{N}_{\bar{k}}\right|$ additional gradients $\overline{\nabla L}_{t,k}(\boldsymbol{\psi}_{t-1,i})$
using neighbor models $\boldsymbol{\psi}_{t-1,i}$ and up to $\left|\mathcal{N}_{\bar{k}}\right|$
additional gradient descent rounds (\ref{eq:adapt-fast}) for local
model update using the neighbor gradients. A quantitative evaluation
of the overhead and the execution time of local computations is proposed
in Sect. \ref{sec:Industrial-IoT:-experiments} by comparing FA, CFA
and CFA-GE using real data and low-power System on Chip (SoC) devices.

Considering now networking aspects, the cost of a D2D communication
is much lower than the cost of a server connection, typically long-range.
D2D links cover shorter ranges and require lower transmit power: communication
cost is thus ruled by the energy spent during receiving operations
(radio wake-up, synchronization, decoding). Besides, in large-scale
and massive IIoT networks, sending model updates to the server, as
done in conventional FA, might need several D2D communication rounds
as relaying information via neighbor devices. D2D communications can
serve as an underlay to the infrastructure (server) network and can
thus exploit the same radio resources. Such two-tier networks are
a key concept in next generation IoT and 5G \cite{magd2d} scenarios.

Finally, optimal trading between in-network and server-side operations
is also possible by alternating rounds of FA with rounds of in-network
consensus (CFA or CFA-GE). This corresponds to a real-world scenario
where communication with the server is available, but intermittent,
sporadic, unreliable or too costly. During initialization, \emph{i.e.}
at time $t=t_{0}$, devices might use the last available global model
received from the server, $\mathbf{W}_{t_{0},k}=\mathbf{W}_{N_{s}}$,
after $N_{s}$ communication rounds of the previous FA phase, and
obtain a local update via SGD: $\mathbf{W}_{t+1,k}=\mathbf{W}_{N_{s}}-\mu_{t}\nabla L_{t,k}(\mathbf{W}_{N_{s}})$.
This is fed-back to neighbors to start CFA or CFA-GE iterations.

\section{Consensus-based FL: an introductory example}

\label{sec:Federated-averaging-by}

\begin{figure}[t!]
\center\includegraphics[scale=0.51]{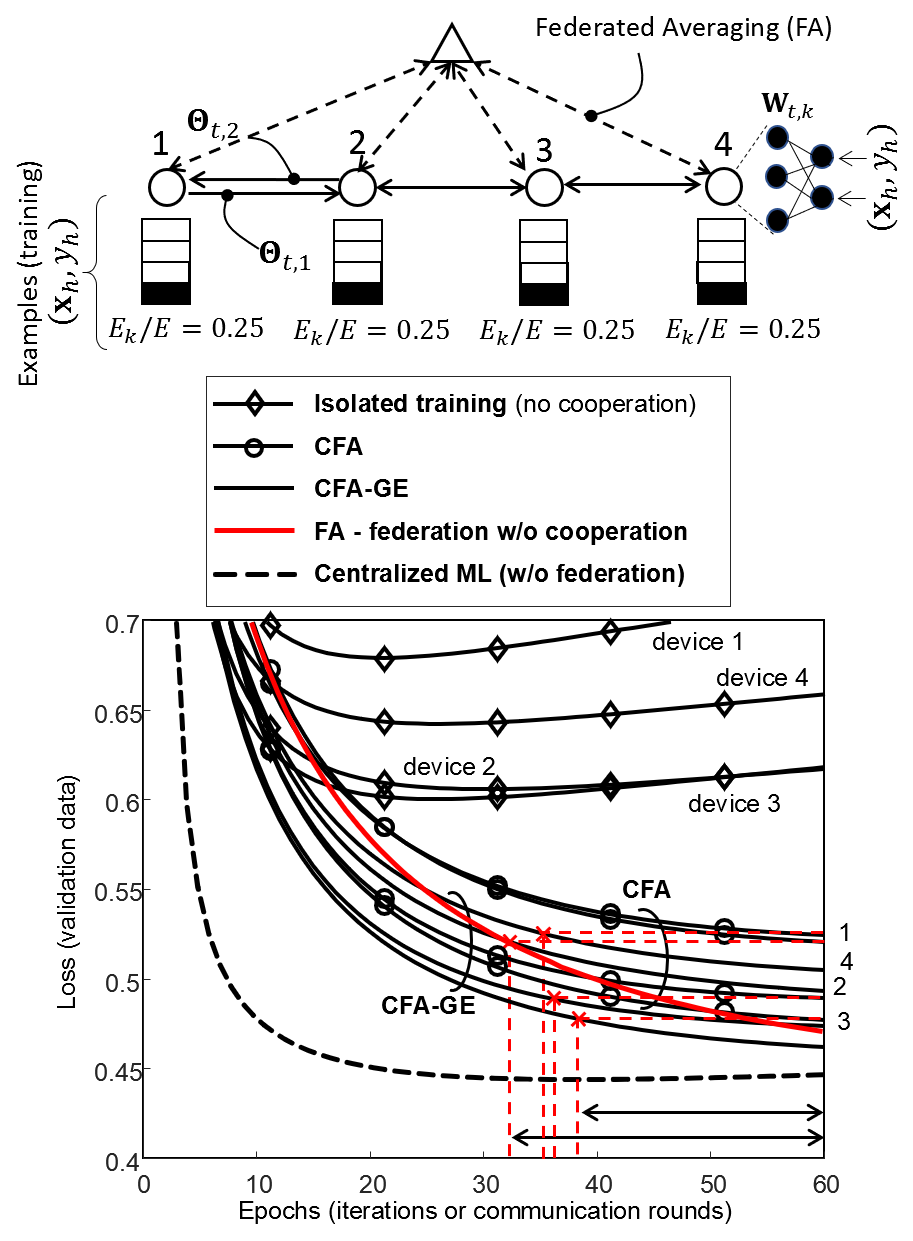} %\par\end{centering}
\protect\caption{\label{intro2} Comparison of FL methods over a multihop wireless
network of $K=4$ devices. Validation loss vs. iterations over the
full MNIST dataset for all devices: CFA (circle markers), CFA-GE (solid
lines without markers), FA (red line), isolated model training (diamond
markers), and centralized ML without federation (dashed line). Iterations
correspond to \emph{epochs} when running inside the server, or \emph{communication
rounds} when running consensus or FA. }
\end{figure}

\begin{figure}[t!]
\center\includegraphics[scale=0.35]{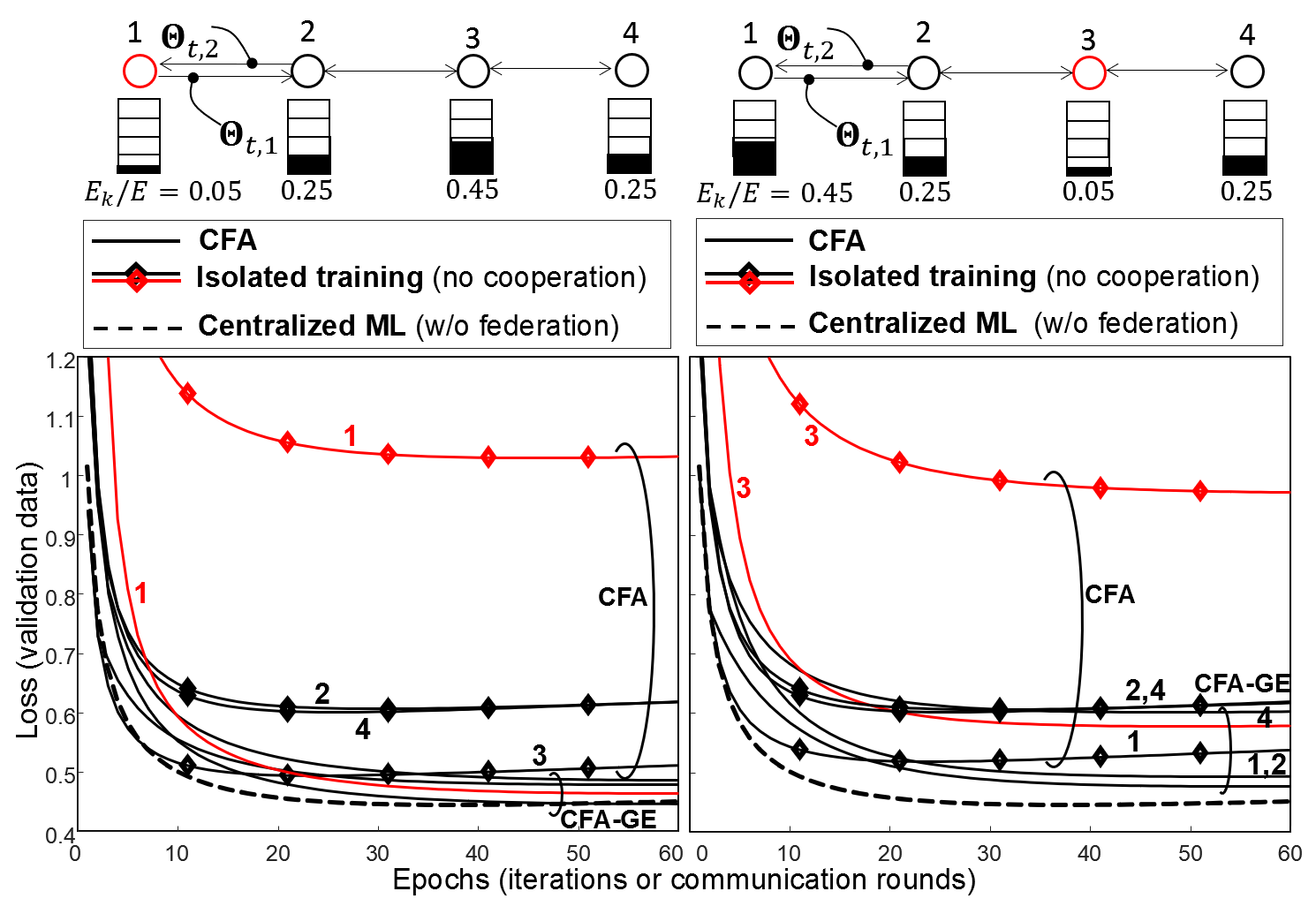} %\par\end{centering}
\protect\caption{\label{UNBALANCED} Effect of non-IID unbalanced training data over
device $k=1$ and $k=3$ (red lines) of a multihop wireless network
composed of $K=4$ devices. Validation loss over the full MNIST dataset
vs. epochs (or consensus iterations). Comparison between CFA (solid
lines), isolated model training (diamond markers), and centralized
ML without federation (dashed line) is also presented. Non-IID data
distribution is shown visually on top, for each case.}
\end{figure}

In this section, we give an introductory example of consensus-based
FL approaches comparing their performance to conventional FL methods.
We resort here to a network of $K=4$ wireless devices communicating
via multihop as depicted in Fig. \ref{intro2} without any central
coordination. Although simple, the proposed topology is still useful
in practice to validate the performance of FL under the assumption
that no device has direct (\ie single-hop) connection with all nodes
in the network. More practical usage scenarios are considered in Sect.
\ref{sec:Industrial-IoT:-experiments}. Without affecting generality,
the devices collaboratively learn a global model $\hat{y}(\mathbf{W};\mathbf{x})$
that is simplified here as a NN model with only one fully connected
layer ($Q=1$): 
\begin{equation}
\hat{y}(\mathbf{W};\mathbf{x})=f_{1}\left(\mathbf{w}_{0,1}^{T}\mathbf{x}+\mathbf{w}_{1,1}\right),\text{\ensuremath{\mathbf{W}}= \ensuremath{\left[\mathbf{w}_{\mathrm{0,1}}^{\mathrm{T}}\mathit{,\mathbf{w}_{\mathrm{1,1}}}\right]}}.\label{eq:simplified}
\end{equation}
Considering the $4$-node network layout, the neighbor sets consist
of $\mathcal{N}_{\bar{1}}=\left\{ 2\right\} $, $\mathcal{N}_{\bar{2}}=\left\{ 1,3\right\} $,
$\mathcal{N}_{\bar{3}}=\left\{ 2,4\right\} $, $\mathcal{N}_{\bar{4}}=\left\{ 3\right\} $.
Each $k$-th device has a database of $E_{k}$ local training data,
$\mathcal{E}_{k}=\{(\mathbf{x}_{h},y_{h})\}_{h=1}^{E_{k}}$, that
are here taken from the MNIST (Modified National Institute of Standards
and Technology) image database \cite{mnist} of handwritten digits.
Output labels $y_{h}$ take $C=10$ different values (from digit $0$
up to $9$), model inputs $\mathbf{x}$ have size $d_{1}=784$ (each
image is represented by $28\times28$ grayscale pixels), while outputs
$\hat{y}$ have dimension $d_{2}=C=10$. In Fig. \ref{intro2}, each
device obtains the same number of training data $(E_{k}=400$ images)
taken randomly (IID) from the database consisting of $E=1600$ images.
Non-IID data distribution is investigated in Fig. \ref{UNBALANCED}.

We assume that each device has prior knowledge of the model (\ref{eq:simplified})
structure at the initial stage ($t=0$), namely the input/output size
($d_{1},d_{2}$) and the non-linear activation $f_{1}(\cdot)$. Moreover,
each of the $K=4$ local models starts from the same random initialization
$\mathbf{W}_{0,k}$ for $t=0$ \cite{mah}. Every new epoch $t>0$,
the devices perform consensus iterations using the model parameters
received from the available neighbors during the previous epoch $t-1$.
Local model updates for CFA (\ref{adapt}) and CFA-GE (\ref{adapt-1})
use the cross-entropy loss for gradient computation 
\begin{equation}
L_{t,k}=-\sum_{h}y_{h}\log\left[\hat{y}(\mathbf{W}_{t,k};\mathbf{x}_{h})\right],\label{crossentr}
\end{equation}
where the sum is computed over mini-batches of size $B=5$. The devices
thus make one training pass over their local dataset consisting of
$E_{k}/B=80$ mini-batches. For CFA, we choose $\epsilon_{t}=1$,
$\mu_{t}=0.025$ and mixing parameters as in (\ref{eq:mix}). For
CFA-GE, the mixing parameters for gradients (\ref{adapt-1-1}) are
selected as $\mu_{t}\beta_{t,k}=0.025$ and $\mu_{t}\beta_{t,i}=0.2$,
$\forall i\in\mathcal{N}_{\bar{k}}$, while the MEWMA hyper-parameter
is set to $\varrho=0.99$.

On every epoch $t$, performance is analyzed in terms of validation
loss (\ref{crossentr}) for all $K=4$ models. For testing, we considered
the full MNIST validation dataset ($\mathbf{x}_{h,val},y_{h,val}$)
consisting of $60.000$ images. The loss $L_{t,k}^{(val)}=-\sum_{h}y_{h,val}\log\left[\hat{y}(\mathbf{W}_{t,k};\mathbf{x}_{h,val})\right]$
decreases over consecutive epochs as far as the model updates $\mathbf{W}_{t,k}$
converge to the \emph{true} global model $\mathbf{W}_{\infty}$.

In Figs. \ref{intro2}-\ref{UNBALANCED}, we validate the performances
of the CFA algorithms in case of uniform (Fig. \ref{intro2}) and
uneven (Fig. \ref{UNBALANCED}) data distribution among the devices.
More specifically, Fig. \ref{intro2} compares CFA and CFA-GE, with
CFA-GE using the two-stage negotiation algorithm of Sect. \ref{subsec:Two-stage-negotiation-and}
and starting\footnote{At initial epochs $t=0,1,2$ we use the $4$-stage negotiation algorithm,
described in Sect. \ref{subsec:Consensus-based-Federated}.} at epoch $t=3$. On the other hand, in Fig. \ref{UNBALANCED}, we
consider the general case where the data is unevenly distributed,
while partitioning among devices is also non IID. Herein, we compare
two cases. In the first one, device $1$ (Fig. \ref{UNBALANCED} on
the left) is located at the edge of the network and connected to one
neighbor only. It obtains $E_{k}=80$ images from only $6$ of the
available $C$ classes, namely the $5$\% of the training database
of $E$ images. Device $3$, connected with $2$ neighbors, retains
a larger database ($E_{k}=720$ images, $45$\% of the training database).
In the second case (on the right), the situation is reversed. As expected,
compared with the first case, convergence is more penalized in the
second case, although CFA running on device $3$ (red lines) can still
converge. As shown in Fig. \ref{intro2}, CFA-GE (solid lines without
markers) further reduces the loss, compared with CFA (circle markers).
Effect of an unbalanced database for CFA-GE is also considered in
Sect. \ref{sec:Industrial-IoT:-experiments}.

\begin{figure*}[!t]
\center\includegraphics[scale=0.36]{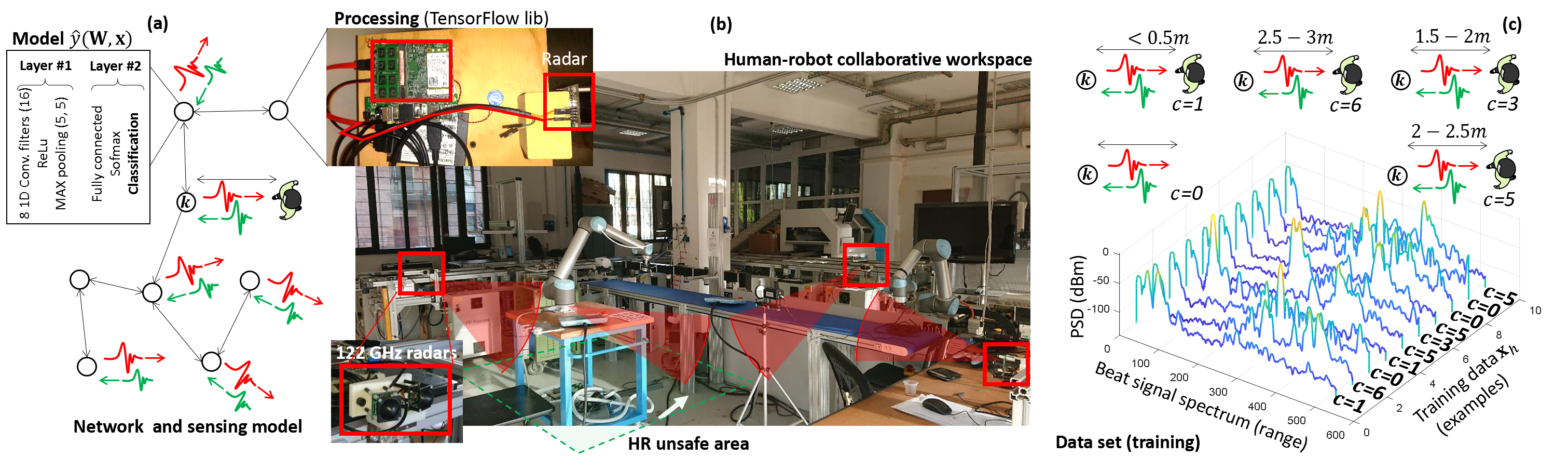} %\par\end{centering}
\protect\caption{\label{thz} From left to right: a) experimental setup: network and
sensing model for in-network federated learning with Convolutional
Neural Network (CNN, top-left corner) examples whose parameters are
shown in Table \ref{tabNNparameters}; b) industrial scenario (CNR-STIIMA
de-manufacturing pilot plant) and deployed radars; c) examples $\mathbf{x}_{h}$
of measured beat signal spectrum ($512$ point FFT) for selected classes
($c=0,1,3,5,6$).}
\end{figure*}

FL and consensus schemes have been implemented using the TensorFlow
library \cite{tf}, while real-time D2D connectivity is simulated
by a MonteCarlo approach. All simulations are running for a maximum
of $60$ epochs. Besides the proposed consensus strategies, validation
loss is also computed for three different scenarios. The first one
is labelled as ``\emph{isolated training}'' and it is evaluated
in Fig. \ref{intro2} for IID and in Fig. \ref{UNBALANCED} for non
IID data. In this scenario, $K=4$ models are trained without any
cooperation from neighbors (or server) by using locally trained data
only. This use case is useful to highlight the benefits of mutual
cooperation that are significant after epoch $t=9$ for IID and after
$t=3$ epochs for non IID, according to the considered network layout.
Notice that isolated training is also limited by overfitting effects
after iteration $t=30$, as clearly observable in Fig. \ref{intro2}
and in Fig. \ref{UNBALANCED}, since local/isolated model optimization
is based only on few training images, compared with the validation
database of $60.000$ images. Consensus and mutual cooperation among
devices thus prevents such overfitting. The second scenario ``\emph{centralized
ML without federation}'' (dashed lines in Figs. \ref{intro2} and
\ref{UNBALANCED}) corresponds to (\ref{nofederation}) and gives
the validation loss obtained when all nodes are sending \emph{all}
their locally trained data directly to the server. It serves as benchmark
for convergence analysis as provides the optimal parameter set $\mathbf{W}$
considering $E=1.600$ images for training. Notice that the CFA-GE
method closely approaches the optimal parameter set and converges
faster than CFA. The third scenario implements the FA strategy (see
Sect. \ref{sec:Federated-optimization-and}) that relies on server
coordination, while cooperation among devices through D2D links is
not enabled. As depicted in Fig. \ref{intro2}, the convergence of
the FA validation loss is similar to those of devices $2$ and $3$,
although convergence speed is slightly faster after epoch $t=55$.
In fact, for the considered network layout, devices $2$ and $3$
can be considered a good replacement of the server, being directly
connected with most of the devices. In the next section, we consider
a more complex device deployment in a IIoT challenging scenario.

\section{Validation in an experimental IIoT scenario}

\label{sec:Industrial-IoT:-experiments}

The proposed in-network FL approaches of Sect. \ref{sec:Federated-averaging-by}
are validated here on a real-world IIoT use case. Data are partitioned
over IIoT devices and D2D connectivity \cite{cloud} is used here
as a replacement to centralized communication infrastructure \cite{iot}.
As depicted in Fig. \ref{thz}, the reference scenario consists of
a large-scale and dense network \cite{cloud0} of autonomous IIoT
devices that are sensing their surroundings using Frequency Modulated
Continuous Wave (FMCW) radars \cite{silicon} working in the $122$
GHz (sub-Thz) band. Radars in the mmWave (or sub-THz) bands are very
effective in industrial production lines (or robotic cells, as in
Fig. \ref{thz}.b) for environment/obstacle detection \cite{passive},
velocity/distance measurement and virtual reality applications \cite{fmcw}.
In addition, mmWave radios have been also considered as candidates
for 5G new radio (NR) allocation. They thus represent promising solutions
towards the convergence of dense communications and advanced sensing
technologies \cite{computer}.

In the proposed setup, the above cited devices are employed to monitor
a shared industrial workspace during Human-Robot Collaboration (HRC)
tasks to detect and track the position of the human operators (\emph{\ie}
the range distance from the individuals) that are moving nearby a
robotic manipulator inside a fenceless space \cite{cellnet}. In industrial
shared workplaces, measuring positions and distance is mandatory to
enforce a worker protection policy, to implement collision avoidance,
reduction of speed, anticipating contacts of limited entity, \emph{etc}.
In addition, it is highly desirable that operators are set free from
wearable devices to generate location information \cite{computer}.
Tracking of body movements must also not depend on active human involvement.
For static background, the problem of passive body detection and ranging
can be solved via background subtraction methods, and ML tools (see
\cite{fmcw2} and references therein). The presence of the robot,
often characterized by a massive metallic size, that moves inside
the shared workplace, poses additional remarkable challenges in ranging
and positioning, because robots induce large, non-stationary, and
fast RF perturbation effects \cite{cellnet}.

The radars collect a large amount of data, that cannot be shipped
back to the server for training and inference, due to the latency
constraints imposed by the worker safety policies. In addition, direct
communication with the server is available but reserved to monitor
the robot activities (and re-planning robotic tasks in case of dangerous
situations) \cite{cellnet} and should not be used for data distribution.
Therefore, to solve the scalability challenge while addressing latency,
reliability and bandwidth efficiency, we let the devices perform model
training without any server coordination but using only mutual cooperation
with neighbors. We thus adopt the proposed in-network FL algorithms
relying solely on local model exchanges over the D2D active links.

In what follows, we first describe the dataset and the ML model $\hat{y}(\mathbf{W};\mathbf{x})$
adopted for body motion recognition. Next, we investigate the convergence
properties of CFA and CFA-GE solutions, namely the required number
of communication rounds (\emph{\ie} latency) for varying connectivity
layouts, network size and hyper-parameters choices, such as mixing
weights and step sizes. Finally, we provide a quantitative evaluation
of the communication overhead and of the local computational complexity
comparing all proposed algorithms.

\subsection{Data collection and processing}

\label{subsec:Data-collection-and}

In the proposed setup, the radar (see \cite{fmcw} for a review) transmitting
antennas radiate a sweeped modulated waveform \cite{silicon} with
bandwidth equal to $6$ GHz, carrier frequency $119$ GHz, and ramp
(pulse) duration set to $T=1$ ms. The radar echoes, reflected by
moving objects are mixed at the receiver with the transmitted signal
to obtain the beat signal. Beat signals are then converted in the
frequency domain (\emph{\ie} beat signal spectrum) by using a $512$-point
Fast Fourier Transform (FFT) and averaged over $10$ consecutive frames
(\emph{\ie} frequency sweeps or ramps). FFT samples are used as model
inputs $\mathbf{x}_{h}$ and serve as training data collected by the
individual devices. The network of radars is designed to discriminate
body movements from robots and, in turn, to detect the distance of
the worker from the robot with the purpose of identifying potential
unsafe conditions \cite{passive}. The ML model is here trained to
classify $C=8$ potential HR collaborative situations characterized
by different HR distances corresponding to safe or unsafe conditions.
In particular, class $c=0$ (model output $\hat{y}=y_{0}$) corresponds
to the robot and the worker cooperating at a safe distance (distance
$\geq3.5$ m), class $c=1$ ($\hat{y}=y_{1}$) identifies the human
operator as working close-by the robot, at distance $<0.5$ m. The
remaining classes are: $c=2$ ($0.5\leq$ distance $<1$ m), $c=3$
($1\leq$ distance $<1.5$ m), $c=4$ ($1.5\leq$ distance $<2$ m),
$c=5$ ($2\leq$ distance $<2.5$ m), $c=6$ ($2.5\leq$ distance
$<3$ m), $c=7$ ($3\leq$ distance $<3.5$ m). The FFT range measurements
(\emph{\ie} beat signal spectrum) and the corresponding true labels
in Fig. \ref{thz}.c, are collected independently by the individual
devices and stored locally. During the initial FL stage, each device
independently obtains $E_{k}=25$ FFT range measurements. Data distribution
is also non-IID: in other words, most of the devices have local examples
only for a (random) subset of the $C=8$ classes of the global model.
However, we assume that there are sufficient examples for each class
considering the data stored on all devices. Local datasets correspond
to the $1$\% of the full training database. Mini-batches for local
gradients have size equal to $B=5$, training passes thus consist
of $E_{k}/B=5$ mini-batches, for fast model update. On the contrary,
validation data consists of $E=16.000$ range measurements collected
inside the industrial plant.

\begin{figure*}[!t]
\center\includegraphics[scale=0.39]{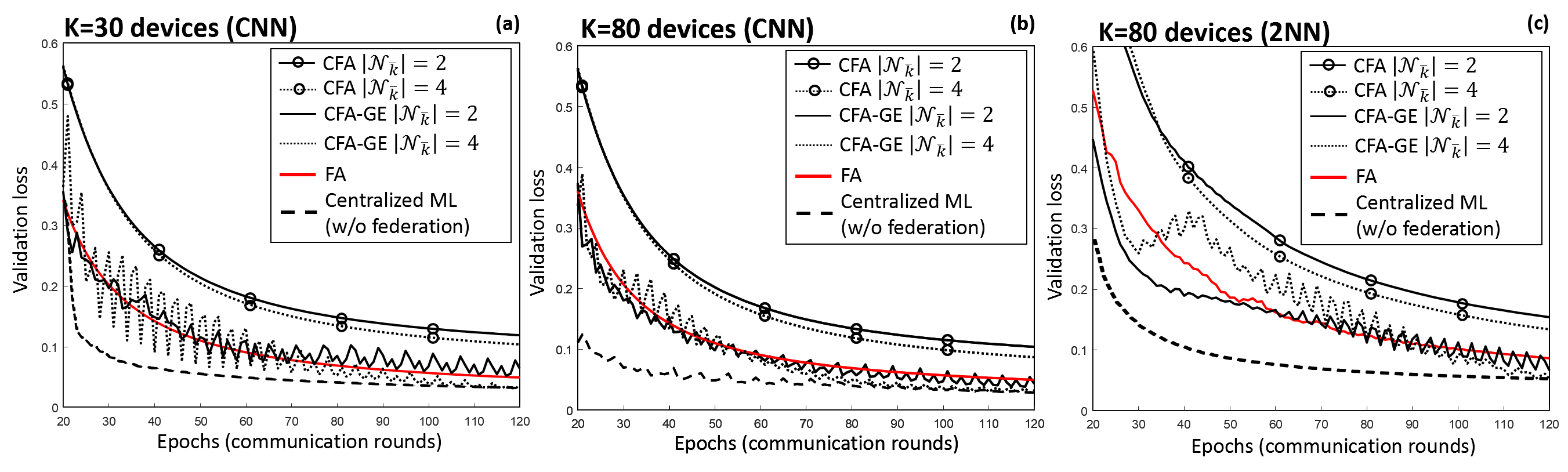} %\par\end{centering}
\protect\caption{\label{thz-results}From left to right: a) FL for CNN model with $K=30$,
b) $K=80$ devices and c) 2NN model with $K=80$ devices. The network
is characterized by $\left|\mathcal{N}_{k}\right|=2$ (solid lines)
and $\left|\mathcal{N}_{k}\right|=4$ (dotted lines) neighbors per
node. Comparative analysis of CFA, CFA-GE, FA (red lines), and centralized
ML \emph{i.e.} learning w/o federation (dashed lines). CNN and 2NN
parameters are described in Table \ref{tabNNparameters}.}
\end{figure*}

\begin{table}[tp]
\begin{centering}
\begin{tabular}{l|l|l|}
 & CNN  & \multicolumn{1}{l}{2-NN}\tabularnewline
\hline 
\multirow{1}{*}{NN model} & \begin{turn}{90}
 $\begin{array}{c}
\textrm{8 1D conv. (16 taps)}\\
\downarrow\\
\textrm{ReLu}\\
\downarrow\\
\textrm{MaxPool (5, 5)}\\
\downarrow\\
\textrm{FC}(168\times C)\\
\downarrow\\
\textrm{Softmax}
\end{array}$ 
\end{turn} & \begin{turn}{90}
 $\begin{array}{c}
\textrm{FC \ensuremath{(512\times32)}}\\
\downarrow\\
\textrm{ReLu}\\
\downarrow\\
\textrm{FC}(32\times C)\\
\downarrow\\
\textrm{Softmax}
\end{array}$ 
\end{turn}\tabularnewline
\hline 
\multirow{2}{*}{Layer $q=1$} & $\mathbf{w}_{0,1}:\begin{array}{l}
d_{1}=16\\
d_{2}=8
\end{array}$  & $\mathbf{w}_{0,1}:\begin{array}{l}
d_{1}=512\\
d_{2}=32
\end{array}$\tabularnewline
\cline{2-3} \cline{3-3} 
 & $\mathbf{w}_{1,1}$: $\begin{array}{c}
d_{1}=8\end{array}$  & $\mathbf{w}_{1,1}$: $\begin{array}{c}
d_{1}=32\end{array}$\tabularnewline
\hline 
\multirow{2}{*}{Layer $q=2$} & $\mathbf{w}_{0,2}$: $\begin{array}{l}
d_{1}=168\\
d_{2}=C
\end{array}$  & $\mathbf{w}_{0,2}$: $\begin{array}{l}
d_{1}=32\\
d_{2}=C
\end{array}$\tabularnewline
\cline{2-3} \cline{3-3} 
 & $\mathbf{w}_{1,2}$: $\begin{array}{c}
d_{1}=C\end{array}$  & $\mathbf{w}_{1,2}$: $\begin{array}{c}
d_{1}=C\end{array}$\tabularnewline
\cline{2-3} \cline{3-3} 
\end{tabular}
\par\end{centering}
\medskip{}
 \protect\caption{\label{tabNNparameters}NN models and trainable parameters $\mathbf{W}$
(weights and biases) for $C=8$ classes.}
\vspace{-0.6cm}
\end{table}

Unlike the previous section, we now choose a ML global model characterized
by a NN with $Q=2$ trainable layers. In particular, two networks
are considered with hyper-parameters and corresponding dimensions
for weights and biases $\mathbf{W}=\left[\mathbf{w}_{0,1}^{\mathrm{T}},\mathbf{w}_{1,1},\mathbf{w}_{0,2}^{\mathrm{T}},\mathbf{w}_{1,2}\right]$
that are detailed in Table \ref{tabNNparameters}. The first convolutional
NN model (CNN) consists of a 1D convolutional layer ($8$ filters
with $16$ taps) followed by max-pooling (non trainable, size $5$
and stride $5$) and a fully connected (FC) layer of dimension $168\times C$.
The second model (2NN) replaces the convolutional layer with an FC
layer of $32$ hidden nodes (dimension $512\times32$) followed by
a ReLu layer and a second FC layer of dimension $32\times C$. The
examples are useful to assess the convergence properties of the proposed
distributed strategies for different layer types, dimensions and number
of trainable parameters. As before, we further assume that, during
the initial stage, each device has knowledge of the ML global model
structure (see layers and dimensions in Table \ref{tabNNparameters}).
At each new communication round, model parameters for each layer are
multiplexed and propagated simultaneously by using a Time Division
Multiple Access (TDMA) scheme \cite{cloud}.

FL has been simulated on a virtual environment but using real data
from the plant. This virtual environment creates an arbitrary number
of virtual devices, each configured to process an assigned training
dataset and exchanging parameters $\Theta_{t,k}$ that are saved in
real-time on temporary cache files. Files may be saved on RAM disks
to speed up the simulation time. The software is written in Python
and uses TensorFlow and multiprocessing modules: simplified configurations
for testing both CFA and CFA-GE setups are also provided in the repository
\cite{ieeedataport}. The code script examples are available as open
source and show the application of CFA and CFA-GE for different NN
models. Hyper-parameters such as learning rates for weights $\alpha_{k,i}$,
gradients $\mu_{t}\beta_{k,i}$, number of neighbors $\mathcal{N}_{\overline{k}}$
and $\varrho$ in (\ref{eq:grad_moving}) are fully configurable.
The data-sets obtained in the scenario of Fig. \ref{thz}.b are also
available in the same repository. Finally, examples have been provided
for implementation and analysis of execution time on low power devices
(Sect. \ref{subsec:Communication-overhead-and-1}). The current optimization
toolkit does not simulate, or account for, packet losses during communication:
this is considered as negligible for short-range connections. However,
the network and connectivity can be time-varying and arbitrarily defined.

\begin{table}[tp]
\begin{centering}
\begin{tabular}{|l|l|l|}
\multicolumn{1}{l|}{Configuration (1)} & Configuration (2)  & \multicolumn{1}{l}{Configuration (3)}\tabularnewline
\hline 
\multirow{2}{*}{$\begin{array}{l}
K<15\\
\left|\mathcal{N}_{\bar{k}}\right|\leq6
\end{array}$ } & \multirow{2}{*}{$\begin{array}{l}
K\in\left[15,50\right]\\
\left|\mathcal{N}_{\bar{k}}\right|\in\left[2,6\right]
\end{array}$} & \multirow{2}{*}{$\begin{array}{l}
K>50\\
\left|\mathcal{N}_{\bar{k}}\right|=2
\end{array}$}\tabularnewline
 &  & \tabularnewline
\hline 
\multicolumn{3}{|c|}{CFA}\tabularnewline
\hline 
$\begin{array}{l}
\epsilon_{t}=1\\
\mu_{t}=0.025\\
\alpha_{k,i}\:\textrm{in }(\ref{eq:mix})
\end{array}$  & $\begin{array}{l}
\epsilon_{t}=0.5\\
\mu_{t}=0.025\\
\alpha_{k,i}\:\textrm{in }(\ref{eq:mix})
\end{array}$  & $\begin{array}{l}
\epsilon_{t}=0.5\\
\mu_{t}=0.025\\
\alpha_{k,i}\:\textrm{in }(\ref{eq:mix})
\end{array}$\tabularnewline
\hline 
\multicolumn{3}{|c|}{CFA-GE}\tabularnewline
\hline 
$\begin{array}{l}
\mu_{t}\beta_{k,i}\leq0.15\\
\varrho=0.99\\
\textrm{(MEWMA)}
\end{array}$  & $\begin{array}{l}
\mu_{t}\beta_{k,i}\in\left[.1,.15\right]\\
\varrho=0.95\div0.99\\
\textrm{(MEWMA)}
\end{array}$  & $\begin{array}{l}
\mu_{t}\beta_{k,i}\leq0.1\\
\varrho=0.9\div0.95\\
\textrm{(MEWMA)}
\end{array}$\tabularnewline
\hline 
\end{tabular}
\par\end{centering}
\medskip{}
 \protect\caption{\label{tab1}Optimized hyper-parameters for CFA and CFA-GE.}
\vspace{-0.6cm}
\end{table}

\subsection{Gradient exchange optimization for NN}

\label{subsec:Gradient-exchange-optimization}

In what follows, FL is verified by varying the number of devices $K=30\div80$
and number of neighbors $\left|\mathcal{N}_{\overline{k}}\right|=2\div10$
to test different D2D connectivity layouts. In Figure \ref{thz-results},
we validate the consensus based FL tool, for both CNN and 2NN models,
over networks with increasing number of devices from $K=30$ to $K=80$.
To simplify the analysis of different connectivity scenarios, the
network is simulated as k-regular (\ie all network devices have the
same number of neighbors, or degree) while we verify realistic topologies
characterized by $\left|\mathcal{N}_{\bar{k}}\right|=2,4,6,10$ neighbors
per node. First, in Fig. \ref{thz-results}, we compare decentralized
CFA and CFA-GE with FA and conventional centralized ML without federation
in (\ref{nofederation}). The chosen optimization hyper-parameters
are summarized in Table \ref{tab1}. For all FL cases (CFA, CFA-GE
and FA), we plot the validation loss vs. communication rounds (\ie
epochs) averaged over all $K$ devices. For centralized ML (dashed
lines), validation loss is analyzed over epochs now running inside
the server. The CFA plots (circle markers) approach slowly the curve
corresponding to FA and centralized ML, while performance improves
in dense networks (dotted lines). CFA-GE curves (solid and dotted
lines without markers) are comparable with FA, and converge after
$50-60$ communication rounds. Use of $\left|\mathcal{N}_{\bar{k}}\right|=2$
neighbors (solid lines) is sufficient to approach FA performances.
Increasing the number of neighbors to $\left|\mathcal{N}_{\bar{k}}\right|=4$
(dotted lines) makes the validation loss comparable with the centralized
ML without federation. Running local SGD on received gradients as
in eq. (\ref{eq:adapt-fast}) causes some fluctuations of the validation
loss as approaching convergence. Fluctuations are due to the (large)
step size $\mu_{t}\beta_{k,i}$ used to combine the gradients every
communication round: learning rate adaptation techniques \cite{sgd}
can be applied for fine tuning. Considering the 2NN model, validation
loss is larger for all cases as the result of the larger number of
model parameters to train, compared with CNN. CFA-GE is still comparable
with FA mostly after $70$ rounds and converges towards centralized
ML after $110$ rounds. In all cases, $\left|\mathcal{N}_{\bar{k}}\right|=2$
neighbors are sufficient to approach FA results. More neighbors provide
performance improvements mostly in small networks ($K=30$ devices),
while it is still useful to match centralized ML performances.

\begin{figure}[!t]
\center\includegraphics[scale=0.46]{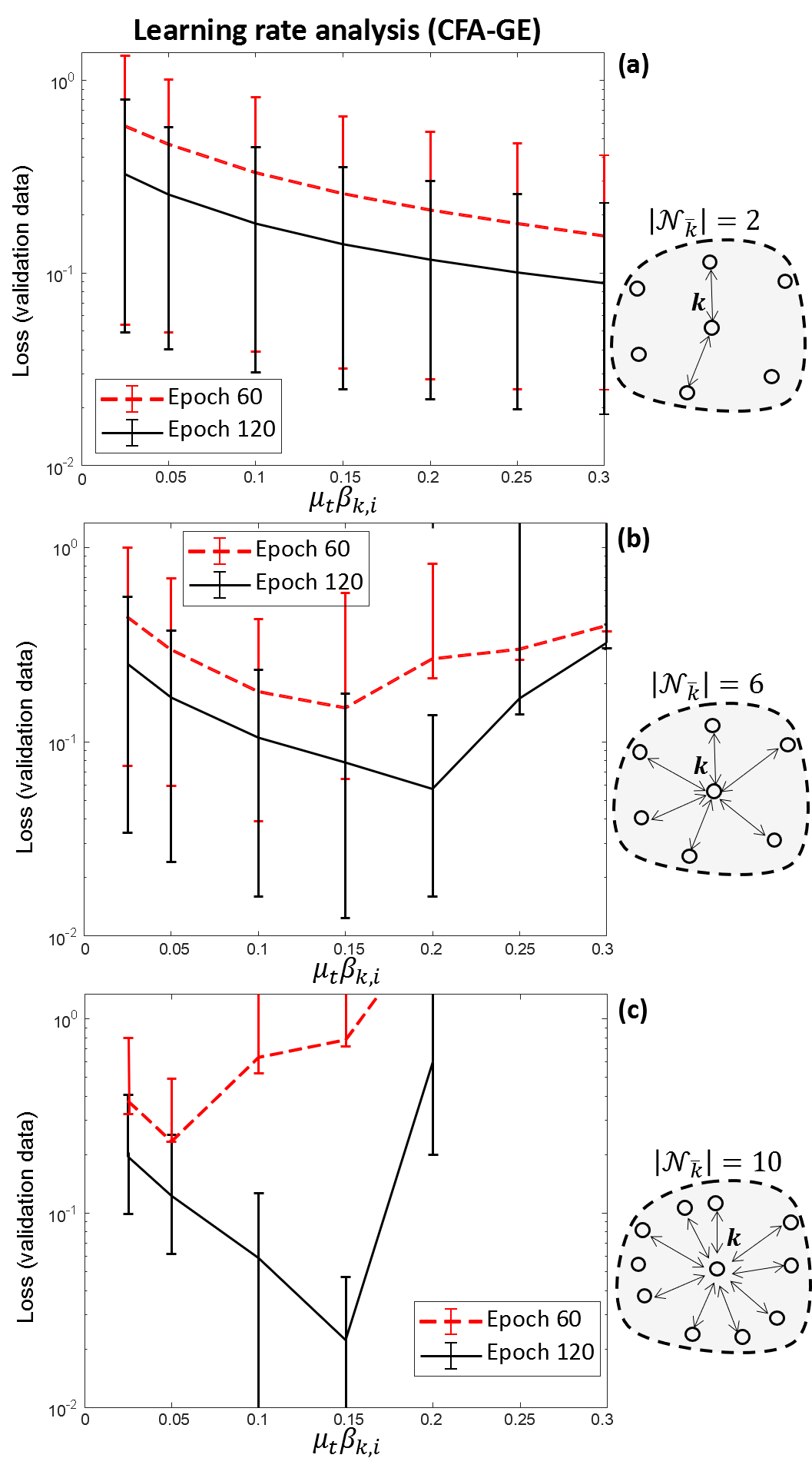} %\par\end{centering}
\protect\caption{\label{thz-results2} From top to bottom: validation loss for varying
rates ($\mu_{t}\beta_{k,i}$) for $K=80$ devices and k-regular networks
with varying connectivity, ranging from a) $\left|\mathcal{N}_{\bar{k}}\right|=2$,
b) $\left|\mathcal{N}_{\bar{k}}\right|=6$ up to c) $\left|\mathcal{N}_{\bar{k}}\right|=10$.}
\end{figure}

In Fig. \ref{thz-results2}, we consider the CFA-GE method and analyze
more deeply the effect of the hyper-parameter choice on convergence,
for $K=80$ devices and varying network degrees $\left|\mathcal{N}_{\bar{k}}\right|$.
The first case ($\left|\mathcal{N}_{\bar{k}}\right|=2$) is representative
of a multihop wireless network; networks with larger degrees ($\left|\mathcal{N}_{\bar{k}}\right|=6$,
$\left|\mathcal{N}_{\bar{k}}\right|=10$) are useful to verify the
performance of FL over denser networks. Line plots with bars in Fig.
\ref{thz-results2} are used to graphically represent the variability
of the validation loss observed by the devices. We analyze the learning
rate ($\mu_{t}\beta_{k,i}$) used to combine the gradients received
from the neighbors in (\ref{eq:adapt-fast}). Other hyper-parameters
are selected as in Table \ref{tab1}. As expected, increasing the
network degree helps convergence and makes the validation loss to
decrease faster since less communication rounds are required. However,
while for $\left|\mathcal{N}_{\bar{k}}\right|=2$ degree networks
the learning rates $\mu_{t}\beta_{k,i}$ can be chosen arbitrarily
in the range $\mu_{t}\beta_{k,i}=0.1\div0.2$ without affecting performance,
denser networks, \emph{i.e.}, with large degree $\left|\mathcal{N}_{\bar{k}}\right|=10$,
require the optimization of the learning rate: smaller rates $\mu_{t}\beta_{k,i}\leq0.1$
improve convergence for $\left|\mathcal{N}_{\bar{k}}\right|=6$ and
$\left|\mathcal{N}_{\bar{k}}\right|=10$.

In Table \ref{learning_layers}, we analyze the latency of the FL
process that is measured here in terms of number of communication
rounds. CFA-GE is considered in detail, while performance of CFA can
be inferred from Fig. \ref{thz-results}. The Table \ref{learning_layers}
reports the number $(t$) of communication rounds (or epochs) that
are required to achieve a target validation loss for all devices,
such that $L_{t,k}^{(val)}\leq0.5$, $\forall k$. For the considered
case, the chosen validation loss of $0.5$ corresponds to a (global)
accuracy of $\gamma_{G}=0.9$. Focusing on CNN layers, a network with
$\left|\mathcal{N}_{\bar{k}}\right|=2$ neighbors per device requires
a max of $21$ communication rounds (and a minimum of $9$) to achieve
a target loss of $L_{t,k}^{(val)}\leq0.5$. This is in line with the
theoretical bound \cite{Ma} $\log\left[\mathrm{1/\left(1-\gamma_{G}\right)}\right]\mathrm{/\gamma_{L}}=15$
for local accuracy $\gamma_{L}\simeq0.2$ (obtained by isolated training).
Considering FA (not shown in the Table), the number of required rounds
ranges from $7$ to $16$, and it is again comparable with decentralized
optimization. Increasing the number of neighbors to $\left|\mathcal{N}_{\bar{k}}\right|=6$,
the required communication rounds reduce to $18$ and to $14$ for
$\left|\mathcal{N}_{\bar{k}}\right|=10$. For 2NN layers, the required
number of epochs increases due to the smaller local accuracy $\gamma_{L}\simeq0.1$
as well as the larger number of parameters to be trained for each
NN layer. Finally, for the proposed setup, we noticed that performance
improves by keeping the learning rate $\mu_{t}\beta_{k,i}$ for the
hidden layer parameters $\left[\mathbf{w}_{0,1}^{\mathrm{T}},\mathbf{w}_{1,1}\right]$,
($q=1$) slightly larger than the rate for the output layer parameters
$\left[\mathbf{w}_{0,2}^{\mathrm{T}},\mathbf{w}_{1,2}\right]$, ($q=2$).
This is particularly evident when convolutional layers are used. 
\begin{table}[tp]
\begin{centering}
\begin{tabular}{c|c|l|l|l|l}
 & Layers  & \multicolumn{2}{c|}{CNN ($L_{t,k}^{(val)}\leq0.5$)} & \multicolumn{2}{c}{2NN ($L_{t,k}^{(val)}\leq0.5$)}\tabularnewline
\hline 
\multirow{2}{*}{$\left|\mathcal{N}_{\bar{k}}\right|$} & \multirow{2}{*}{$(\mathbf{W})$} & \multirow{2}{*}{$\mu_{t}\beta_{k,i}$ } & Epochs ($t$)  & \multirow{2}{*}{$\mu_{t}\beta_{k,i}$ } & Epochs ($t$)\tabularnewline
 &  &  & \multicolumn{1}{l|}{(min | max)} &  & (min | max)\tabularnewline
\hline 
\multirow{4}{*}{$2$} & $q=1$  & $\mathbf{0.2}$  & \multirow{2}{*}{$9$ | $\mathbf{21}$} & $0.05$  & \multirow{2}{*}{$11$ | $51$}\tabularnewline
 & $q=2$  & $\mathbf{0.15}$  &  & $0.1$  & \tabularnewline
\cline{2-6} \cline{3-6} \cline{4-6} \cline{5-6} \cline{6-6} 
 & \emph{\noun{$q=1$}}  & \emph{\noun{$0.15$}}  & \multirow{2}{*}{$10$ | $26$} & \emph{\noun{$\mathbf{0.1}$}}  & \multirow{2}{*}{$12$ | $\mathbf{23}$}\tabularnewline
 & \emph{\noun{$q=2$}}  & \emph{\noun{$0.2$}}  &  & \emph{\noun{$\mathbf{0.05}$}}  & \tabularnewline
\hline 
\multirow{4}{*}{$6$ } & $q=1$  & $\mathbf{0.15}$  & \multirow{2}{*}{$8$ | $\mathbf{18}$} & $0.025$  & \multirow{2}{*}{$17$ | $28$}\tabularnewline
 & $q=2$  & $\mathbf{0.1}$  &  & $0.05$  & \tabularnewline
\cline{2-6} \cline{3-6} \cline{4-6} \cline{5-6} \cline{6-6} 
 & $q=1$  & $0.1$  & \multirow{2}{*}{$12$ | $23$} & $\mathbf{0.05}$  & \multirow{2}{*}{$11$ | $\mathbf{19}$}\tabularnewline
 & $q=2$  & $0.15$  &  & $\mathbf{0.025}$  & \tabularnewline
\hline 
\multirow{4}{*}{$10$ } & $q=1$  & $\mathbf{0.05}$  & \multirow{2}{*}{$7$ | $\mathbf{14}$} & $0.025$  & \multirow{2}{*}{$15$ | $23$}\tabularnewline
 & $q=2$  & $\mathbf{0.025}$  &  & $0.02$5  & \tabularnewline
\cline{2-6} \cline{3-6} \cline{4-6} \cline{5-6} \cline{6-6} 
 & $q=1$  & $0.025$  & \multirow{2}{*}{$11$ | $\mathbf{17}$} & $\mathbf{0.05}$  & \multirow{2}{*}{$10$ | $\mathbf{17}$}\tabularnewline
 & $q=2$  & $0.05$  &  & $\mathbf{0.025}$  & \tabularnewline
\cline{2-3} \cline{3-3} \cline{5-5} 
\end{tabular}
\par\end{centering}
\medskip{}
 \protect\caption{\label{learning_layers}Communication rounds (epochs $t$) for target
validation loss: $L_{t,k}^{(val)}\protect\leq0.5$, $\forall k$.
Min. and max. epochs over $K=80$ devices for CNN and 2NN. Optimized
parameters (bold) are shown, too.}
\vspace{-0.6cm}
\end{table}

\subsection{Communication and computational cost assessment}

\label{subsec:Communication-overhead-and-1}

The CFA-GE method achieves the performance of the centralized ML without
federation, in exchange for a more intensive use of D2D wireless links
and local computations, that scale in both cases with the number of
neighbors $\left|\mathcal{N}_{\bar{k}}\right|$. Based on the the
analysis in Sect. \ref{subsec:Communication-overhead-and}, in Table
\ref{cost} we compare the communication overhead and computational
cost for varying number of neighbors $\left|\mathcal{N}_{k}\right|$,
considering FA, CFA and CFA-GE methods. In particular, the communication
overhead quantifies the number of bytes that need to be transmitted
over-the-air by each device every communication/consensus round. The
computational cost is measured in terms of average local execution
time per communication round and device.

The communication overhead of FA and CFA is $2.98$ Kbyte/round/device
for CNN and $33.36$ Kbyte/round/device for 2NN. Overhead corresponds
in both cases to the model $\mathbf{W}$ size: this is evaluated according
to the model parameters highlighted in Table \ref{tabNNparameters}
and assuming $16$ bit/parameter quantization. CFA-GE overhead is
larger as it scales linearly with $\left|\mathcal{N}_{k}\right|$,
therefore it can be quantified as $2.98\cdot\left|\mathcal{N}_{k}\right|$
Kbyte/round/device for CNN and $33.36\cdot\left|\mathcal{N}_{k}\right|$
Kbyte/round/device for 2NN. Notice that bandwidth-limited communication
systems, \eg based on IEEE 802.15.4, 6LoWPAN and related evolutions
\cite{cloud}-\cite{iot}, are characterized by small physical frame
payloads, typically below $1$ Kbyte/frame. Therefore, sending FA,
CFA or CFA-GE parameters on each round might require the aggregation
of consecutive physical frames, or multiple network layer transactions.
For comparison with centralized ML, FFT training measurements collected
individually by devices have size within the range $1\div4$ Mbyte,
assuming $32$ bit quantization for in-phase (I) and quadrature (Q)
components.

Local execution time considers here the ML stages only, while data
pre-processing and acquisition steps are not included, being negligible
compared to the learning steps. The execution time shown in Table
\ref{cost} is measured using the \emph{timeit} Python module, on
a device equipped with a $1.5$ GHz quad core ARM Cortex-A72 processor
with 4 GB internal RAM\footnote{Typical commercial low-power single-board computer (Raspberry Pi 4
Model B): smaller execution times are expected when running on dedicated
TPU processors}. Focusing on a realistic IIoT environment, the device has thus limited
computational capabilities, compared with the server. Execution time
depends in general on the specific CPU or Tensor Processing Unit (TPU)
performances; nevertheless, the analysis of the results in Table \ref{cost}
is useful to highlight the scaling performance of the proposed methods
compared to the plain FA algorithm. Considering FA, the execution
time on the device is ruled by local SGD rounds: it is $140$ms and
$145$ms for CNN and 2NN, respectively. Notice that FA needs the server
for aggregation, while such additional cost is not considered here.
Compared with FA, CFA adds a model aggregation stage (\ref{eq:aggreg})
for each NN layer that takes $0.5$ ms on average per neighbor. Finally,
CFA-GE adds a cumulative time of $90$ ms for CNN and $94$ ms for
2NN on average per neighbor. This is needed for the computation of
one additional gradient, the MEWMA update (\ref{eq:grad_moving}),
and one SGD round (\ref{eq:adapt-fast}) per neighbor.

Considering both overhead and local execution time, CFA-GE cost per
round is higher compared with CFA and FA and scales almost linearly
with the number of neighbors $\left|\mathcal{N}_{\bar{k}}\right|$.
CFA total cost is instead comparable with FA when $\left|\mathcal{N}_{\bar{k}}\right|<10$.
However, it is worth to notice that, as shown by the numerical results
in Sect. \ref{subsec:Gradient-exchange-optimization}, CFA-GE only
needs $\left|\mathcal{N}_{\bar{k}}\right|=2$ neighbors for convergence
and even with such a low degree of cooperation it reduces the number
of rounds by almost one order of magnitude (Fig. \ref{thz-results})
compared to CFA. Using more than $\left|\mathcal{N}_{\bar{k}}\right|=2$
cooperating neighbors for CFA-GE provides only marginal improvements
and it is thus not recommended. Having said that, we can conclude
that CFA-GE is promising as an effective replacement for FA and centralized
ML when it is critical to limit the number of communication rounds,
or in case frequent learning updates are needed. CFA keeps complexity
and overhead comparable with FA in exchange for more rounds. It is
thus suitable for non-critical learning tasks and it can support bandwidth-limited
D2D communication systems characterized by small frame payloads.

\begin{table}[tp]
\begin{centering}
\begin{tabular}{c|l|l|l|l|l}
 & \multicolumn{3}{c|}{Comm. overhead } & \multicolumn{2}{c}{Execution time (avg.)}\tabularnewline
 & \multicolumn{3}{c|}{{[}Kbyte/round/device{]}} & \multicolumn{2}{c}{{[}m.sec./round/device{]}}\tabularnewline
\hline 
$\left|\mathcal{N}_{\bar{k}}\right|$  & \multicolumn{1}{l|}{} & CNN  & 2NN  & CNN  & 2NN\tabularnewline
\hline 
\multirow{3}{*}{$2$} & FA  & $2.98$  & $33.36$  & $140$ms  & $145$ms\tabularnewline
 & CFA  & $2.98$  & $33.36$  & $141$ms  & $146$ms\tabularnewline
 & CFA-GE  & $5.96$  & $66.72$  & $321$ms  & $334$ms\tabularnewline
\hline 
\multirow{3}{*}{$6$} & FA  & $2.98$  & $33.36$  & $140$ms  & $145$ms\tabularnewline
 & CFA  & $2.98$  & $33.36$  & $142$ms  & $147$ms\tabularnewline
 & CFA-GE  & $17.88$  & $200.16$  & $684$ms  & $711$ms\tabularnewline
\hline 
\multirow{3}{*}{$10$} & FA  & $2.98$  & $33.36$  & $140$ms  & $145$ms\tabularnewline
 & CFA  & $2.98$  & $33.36$  & $149$ms  & $153$ms\tabularnewline
 & CFA-GE  & $29.8$  & $333.7$  & $984$ms.  & $\sim1$sec.\tabularnewline
\end{tabular}
\par\end{centering}
\medskip{}
 \protect\caption{\label{cost}Communication overhead {[}Kbyte/round/device{]} and computational
cost - average execution time {[}m.sec./round/device{]} measured on
the device.}
\vspace{-0.6cm}
 
\end{table}

\section{Conclusions and open problems}

\label{sec:Conclusions-and-open}

The paper addressed a new family of FL methods that leverage the mutual
cooperation of devices in distributed wireless IoT networks without
relying on the support of a central coordinator. The adaptation of
federated averaging, namely the CFA method, was first discussed to
exploit distributed consensus paradigms for FL. Next, to improve convergence
speed, we proposed a new algorithm based on iterative exchange of
model updates and gradients. The CFA-GE algorithm was optimized for
two-stage gradient negotiations so that it can be tailored to arbitrarily
large scale networks.

The proposed distributed learning approach was validated on an IIoT
scenario where a NN model was distributedly trained to solve the problem
of passive body detection inside a human-robot collaborative workspace.
CFA-GE was shown to achieve the performance of server-side (or centralized)
federated optimization. Decentralized optimization is fully server-less
with respect to NN model training as intelligence is pushed down into
the IoT devices. This is particularly effective when direct communication
with the infrastructure is reserved for critical tasks (\emph{\eg}
to control the robot or to perform safety tasks). Motivated by the
growing range of ML applications that will be deployed in 5G and beyond
wireless networks, FL via consensus emerges in this paper as a promising
framework for flexible model optimization over networks characterized
by decentralized connectivity patterns as in massive IoT implementations.

Despite the promising features, the proposed consensus based approach
is giving rise to new challenges, which need to be taken into account
during the system deployment. For example, the paper considered a
simple enough NN model for optimization running on IoT devices with
limited computation capabilities. However, training of deeper networks
on constrained devices is expected to become the mainstream in the
near future \cite{squeeze}. Application of consensus based federated
optimization to deeper networks might require a more efficient use
of the limited bandwidth, including quantization, compression or ad-hoc
channel encoding \cite{ml3}. As revealed in the considered case study,
tweaking of the model hyper-parameters \cite{distillation} as well
as optimizing the learning rates for each NN model layer separately
are also viable solutions to limit the number of communication rounds.
Finally, although the paper addressed a classification task as application
of federated optimization, both CFA and CFA-GE can be generalized
to perform a wider range of computations.

\section*{Appendix}

\subsection*{Appendix A: Comparing CFA and CFA-GE\label{sec:Appendix:-Comparing-federated}}

Combining (\ref{adapt-1-1}) and (\ref{adapt-1}), we obtain the update
equation for CFA-GE

\begin{equation}
\begin{array}{c}
\mathbf{W}_{t+1,k}=\underset{\widetilde{\boldsymbol{\psi}}_{t,k}}{\underbrace{\boldsymbol{\psi}_{t,k}-\mu_{t}\sum_{i\in\mathcal{N}_{\bar{k}}}\beta_{t,i}\nabla L_{t,i}\left(\boldsymbol{\psi}_{t,k}\right)}-}\\
-\mu_{t}\beta_{t,k}\left[\nabla L_{t,k}\left(\widetilde{\boldsymbol{\psi}}_{t,k}\right)\right]
\end{array}\label{eq:ge}
\end{equation}
with $\boldsymbol{\psi}_{t,k}$ in (\ref{eq:aggreg}). This is comparable
with the CFA approach in (\ref{adapt}). Unlike CFA, gradient exchange
terms $\sum_{i\in\mathcal{N}_{\bar{k}}}\beta_{t,i}\nabla L_{t,i}\left(\boldsymbol{\psi}_{t,k}\right)$
allow to incorporate the influence of training data collected by the
neighborhood of device $k$ through the gradient terms $L_{t,i}\left(\boldsymbol{\psi}_{t,k}\right)$.
Computation of $\widetilde{\boldsymbol{\psi}}_{t,k}$ (\ref{adapt-1-1})
thus corresponds to an initial gradient descent round using the training
data from the neighborhood, while subsequent rounds (\ref{eq:ge})
are performed using SGD over local data mini-batches.

\subsection*{Appendix B: Application to SGD with momentum}

SGD is considered throughout the paper as a popular optimization strategy;
however it shows slow convergence properties in some ML problems \cite{sgd}.
Recently, the method of momentum inspired several algorithms (such
as RMSProp and Adam \cite{adam}) optimized for accelerated learning.
Considering that in federated optimization any improvement in the
convergence speed is beneficial in terms of bandwidth usage and latency,
we address the necessary adaptations of CFA-GE strategy to leverage
momentum information. With respect to SGD, momentum addresses the
problem of imperfect estimation of the stochastic gradients as well
as the conditioning of the Hessian matrix \cite{momentum}. Poor estimation
of gradients is even more critical when considering distributed learning
setups. Compared with SGD, the use of momentum modifies the local
update rule (\ref{adapt}),(\ref{adapt-1}) as 
\begin{equation}
\begin{cases}
\begin{array}{c}
\mathbf{W}_{t+1,k}=\boldsymbol{\psi}_{t,k}+\mathbf{\nu}_{t+1,k}\\
\mathbf{\nu}_{t+1,k}=\varrho\mathbf{\nu}_{t,k}-\mu_{t}\nabla L_{t,k}(\boldsymbol{\psi}_{t,k})
\end{array}\end{cases}\label{momentum-1}
\end{equation}
where both current and past gradients contribute to the local model
update (for device $k)$ as multiplied by an exponentially decaying
function ruled by hyper-parameter $\varrho\in\left[0,1\right)$. $\mathbf{\nu}_{t,k}$
is the momentum, or velocity, of the gradient descent particle at
round $t$, that is stored by device $k$.

Momentum based techniques can be used seamlessly combined with CFA
by simply replacing (\ref{adapt}) with (\ref{momentum-1}) as gradient
sharing is not permitted. Considering that in CFA-GE local gradients
are exchanged among neighbors, some necessary adaptations are required
to leverage momentum. The received gradients, \emph{i.e.} at epoch
$t-1$, are now used for local momentum update as

\begin{equation}
\mathbf{\nu}_{t,k}=\varrho\mathbf{\nu}_{t-1,k}-\mu_{t}\sum_{i\in\mathcal{N}_{\bar{k}}}\beta_{k,i}\mathcal{P}_{\Theta}\left[\overline{\nabla L}_{t-1,i}\right].\label{momentum_adapt-1}
\end{equation}
We then replace (\ref{eq:adapt-fast}) with $\widetilde{\boldsymbol{\psi}}_{t,k}=\boldsymbol{\psi}_{t,k}+\mathbf{\nu}_{t,k}$
and (\ref{adapt-1}) with (\ref{momentum-1}). The use of the Nesterov
momentum \cite{momentum} policy requires minor adaptations: the devices
should now exchange the parameters $\Theta_{t,k}:=\left[\boldsymbol{\psi}_{t,k}+\varrho\mathbf{\nu}_{t-1,k},\nabla\mathbf{L}_{t,k}\right].$
These are: the local model $\boldsymbol{\psi}_{t,k}+\varrho\mathbf{\nu}_{t-1,k}$
\emph{after} applying the velocity term (\ref{momentum_adapt-1})
and the gradients $\nabla\mathbf{L}_{t,k}:=\left\{ \overline{\nabla L}_{t,k}(\mathbf{\boldsymbol{\psi}}_{t-1,i}),\text{}\forall i\in\mathcal{N}_{\bar{k}}\right\} $
described as 
\begin{eqnarray}
\begin{array}{cccc}
\overline{\nabla L}_{t,k}(\boldsymbol{\psi}_{t-1,i}) & = & \varrho\nabla L_{t,k}(\boldsymbol{\psi}_{t-1,i}+\varrho\mathbf{\nu}_{t-1,i})+\\
 &  & +(1-\varrho)\overline{\nabla L}_{t-1,k},
\end{array}\label{eq:grad_moving-1}
\end{eqnarray}
that replace the terms in eq. (\ref{eq:grad_moving}).

\subsection*{Appendix C: Description of Python scripts and datasets}

\setlength{\parindent}{0cm} The section describes the database that
contains the range measurements obtained from FMCW THz radars inside
the Human-Robot (HR) workspace. It also gives examples about how to
use the sample Python code that implements the federated learning
stages via consensus. The database is located in the folder $\mathrm{dati_{-}radar_{-}05-07-2019}$
and can be easily imported through Python:

\begin{lstlisting}
import scipy.io as sio
db = sio.loadmat('dati_radar_05-07-2019
/data_base_all_sequences_random.mat')
\end{lstlisting}

The database contains 5 files:

i) Data\_test\_2.mat has dimension 16000 x 512. Contains 16000 FFT
range measurements (512-point FFT of beat signal after DC removal)
used for test. The corresponding labels are in label\_test\_2.mat
\begin{lstlisting}
x_val = db('Data_test_2')
x_val_l = db('label_test_2')
\end{lstlisting}

ii) Data\_train\_2.mat has dimension 16000 x 512. Contains 16000 FFT
range measurements (512-point FFT of beat signal after DC removal)
used for training. The corresponding labels are in lable\_train\_2.mat
\begin{lstlisting}
x_t = db('Data_train_2')
x_l = db('label_train_2')
\end{lstlisting}

iii) label\_test\_2.mat with dimension 16000 x 1, contains the true
labels for test data (Data\_test\_2.mat), namely classes (true labels)
correspond to integers from 0 to 7. Class 0: human worker at safe
distance >3.5m from the radar (safe distance) Class 1: human worker
at distance (critical) <0.5m from the corresponding radar Class 2:
human worker at distance (critical) 0.5m - 1m from the corresponding
radar Class 3: human worker at distance (critical) 1m - 1.5m from
the corresponding radar Class 4: human worker at distance (safe) 1.5m
- 2m from the corresponding radar Class 5: human worker at distance
(safe) 2m - 2.5m from the corresponding radar Class 6: human worker
at distance (safe) 2.5m - 3m from the corresponding radar Class 7:
human worker at distance (safe) 3m - 3.5m from the corresponding radar.

iv) label\_train\_2.mat with dimension 16000 x 1, contains the true
labels for train data (Data\_train\_2.mat), namely classes (true labels)
correspond to integers from 0 to 7. See item (iii) for class descriptions.

v) permut.mat (1 x 16000) contains the chosen random permutation for
data partition among nodes/device and federated learnig simulation
(see python code) 

\subsection*{Python code}

Usage example for federated\_sample\_XXX\_YYY.py.
\begin{itemize}
\item  XXX refers to the ML model. Available options: CNN, 2NN
\item  YYY refers to the consensus-based federated learning method. Available options: CFA, CFA-GE.\bigskip{}
\end{itemize}
Run\bigskip{}

python federated\_sample\_XXX\_YYY.py -h\bigskip{}

for help. CFA and CFA-GE are described in Sect. III.A, and Sect. III.B, respectively. The code implements the two-stage implementation of CFA-GE (Sect. III.C).\bigskip{}

For CNN network and CFA-GE use:
federated\_sample\_CNN\_CFA-GE.py {[}-h{]} {[}-l1 L1{]} {[}-l2 L2{]}
{[}-mu MU{]} {[}-eps EPS{]} {[}-K K{]} {[}-N N{]} {[}-T T{]} {[}-ro
RO{]}\bigskip{}

For 2-NN network and CFA-GE use:
federated\_sample\_2NN\_CFA-GE.py {[}-h{]} {[}-l1 L1{]} {[}-l2 L2{]}
{[}-mu MU{]} {[}-eps EPS{]} {[}-K K{]} {[}-N N{]} {[}-T T{]} {[}-ro
RO{]}\bigskip{}
Arguments:

-l1: l1 sets the learning rate (gradient exchange) for convolutional
layer

-l2: l2 sets the learning rate (gradient exchange) for FC layer

-mu: mu sets the learning rate for local SGD

-eps: eps sets the mixing parameters for model averaging (CFA)

-K: K sets the number of network devices

-N: N sets the number of neighbors per device

-T T sets the number of training epochs

-ro ro sets the hyperparameter for MEWMA
\bigskip{}

Optional arguments:

-h, -{}-help show this help message and exit.\bigskip{}

For testing CFA performance with CNN network, please use:

federated\_sample\_CNN\_CFA.py [-h] [-mu MU] [-eps EPS] [-K K] [-N N] [-T T]\bigskip{}

Similarly, for 2-NN network, please use:

federated\_sample\_2NN\_CFA.py [-h] [-mu MU] [-eps EPS] [-K K] [-N N] [-T T]\bigskip{}

Optional arguments:

-h, --help show this help message and exit

-mu MU sets the learning rate for local SGD

-eps EPS sets the mixing parameters for model averaging (CFA)

-K K sets the number of network devices

-N N sets the number of neighbors per device

-T T sets the number of training epochs\bigskip{}

Example 1:

\begin{lstlisting}
python federated_sample_CNN_CFA-GE.py 
-l1 0.025 -l2 0.02 -K 40 -N 2  
-T 40 -ro 0.99 
\end{lstlisting}
Use convolutional layers followed by a FC layer (see Table I, CNN
network). Sets gradient learning rate for hidden layer to 0.025, for
output layer to 0.02, K=40 devices, N=2 neighbors per device, MEWMA
parameter 0.99 (see Sect. III).\bigskip{}

Example 2:

\begin{lstlisting}
python federated_sample_2NN_CFA-GE.py 
-l1 0.01 -l2 0.015 -N 2 -T 40 -ro 0.99
\end{lstlisting}
Use FC layers only (see Table I, 2-NN network). Sets gradient learning
rate for hidden layer to 0.01, for output layer to 0.015, K=80 devices
(default), N=2 neighbors per device, MEWMA parameter 0.99.

\subsection*{Python package description}
Python package can be downloaded also from  https://test.pypi.org/project/consensus-stefano/0.3/. 

To initialize CFA use the constructor: 

consensus\_p = CFA\_process(federated, tot\_devices, device\_id, neighbors\_number). 

Use similar constructor for CFA-GE: 

consensus\_p = CFA-ge\_process(federated, tot\_devices, device\_id, neighbors\_number, ro).\bigskip{}

Example:

\begin{lstlisting}
consensus_p = CFA_process(True, 80, 2, 2)
\end{lstlisting}
Initialize CFA process on device 2 with 2 neighbors and for a network of 80 devices.\bigskip{} 

To apply/update federated weights use: consensus\_p.getFederatedWeight( ... )\bigskip{}

To enable/disable consensus (dynamically) consensus\_p.disable\_consensus( ... True/False ... )

\subsection*{Changing ML network parameters}
ML network parameters are defined in the Python scripts federated\_sample\_XXX\_YYY.py. In particular considering CFA-GE script, the CNN network is defined at lines 99-109 by
\begin{lstlisting}
# Construct model 
# Layer #1 CNN 1d, 
# Layer #2 FC
hidden0 = conv1d(x, W_ext_l1, b_ext_l1)
hidden01 = tf.layers.max_pooling1d(
hidden0, pool_size=stride, 
strides=stride, padding='SAME')
fc01 = tf.reshape(hidden01, 
[-1, multip*number])
pred = tf.nn.softmax(
tf.matmul(fc01, W_ext_l2) + b_ext_l2)  
\end{lstlisting}
with parameters defined in lines 39-45. Modify these lines to change the CNN network, namely changing filter and stride sizes, or adding further layers.

\end{document}